\documentclass[
12pt, 
english, 
singlespacing, 
headsepline, 
]{MastersDoctoralThesis} 

\usepackage[utf8]{inputenc} 
\usepackage[T1]{fontenc} 

\usepackage{palatino} 
\usepackage{subcaption}
\usepackage{textcomp}
\usepackage{amsmath}

\usepackage[backend=biber,style=numeric,natbib=true]{biblatex} 

\usepackage[autostyle=true]{csquotes} 

\thesistitle{Merging real and virtual worlds: An analysis of the state of the art and practical evaluation of Microsoft Hololens}
\supervisor{Prof. Tom \textsc{Mens}} 
\supervisoremail{tom.mens@umons.ac.be}
\examiner{} 
\degree{Master in Computer Science} 
\author{Adrien \textsc{Coppens}} 
\authoremail{adrien.coppens@student.umons.ac.be}
\addresses{} 

\subject{Computer Science} 
\university{\href{http://www.umons.ac.be}{University of Mons}} 
\department{\href{http://www.informatique.umons.ac.be}{Department of Computing Science}} 
\group{\href{http://informatique.umons.ac.be/genlog}{Software Engineering Lab}} 
\faculty{\href{http://portail.umons.ac.be/en2/universite/facultes/fs/}{Faculty of Sciences}} 

\AtBeginDocument{
\hypersetup{pdftitle=\ttitle} 
\hypersetup{pdfauthor=\authorname} 
\hypersetup{pdfkeywords=\keywordnames, colorlinks=false} 
}

\newcommand{\blankpage}{
	\newpage
	\thispagestyle{empty}
	\mbox{}
	\newpage
}

\begin{document}

\frontmatter 

\pagestyle{plain} 


\begin{titlepage}
	\vspace*{0.5cm}
\begin{center}
\begin{figure}[!htb]
	\begin{minipage}{0.48\textwidth}
		\centering
		\includegraphics[width=.7\linewidth]{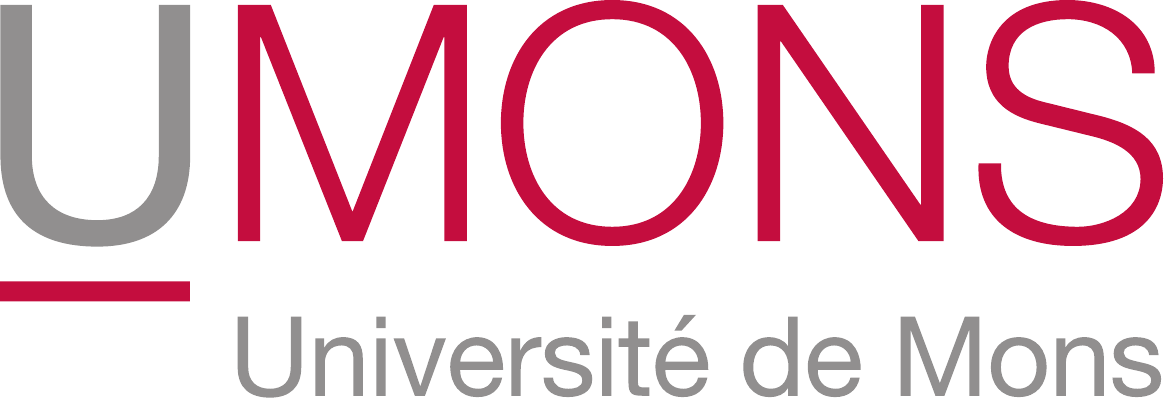}
	\end{minipage}\hfill
	\begin {minipage}{0.48\textwidth}
	\centering
	\includegraphics[width=.65\linewidth]{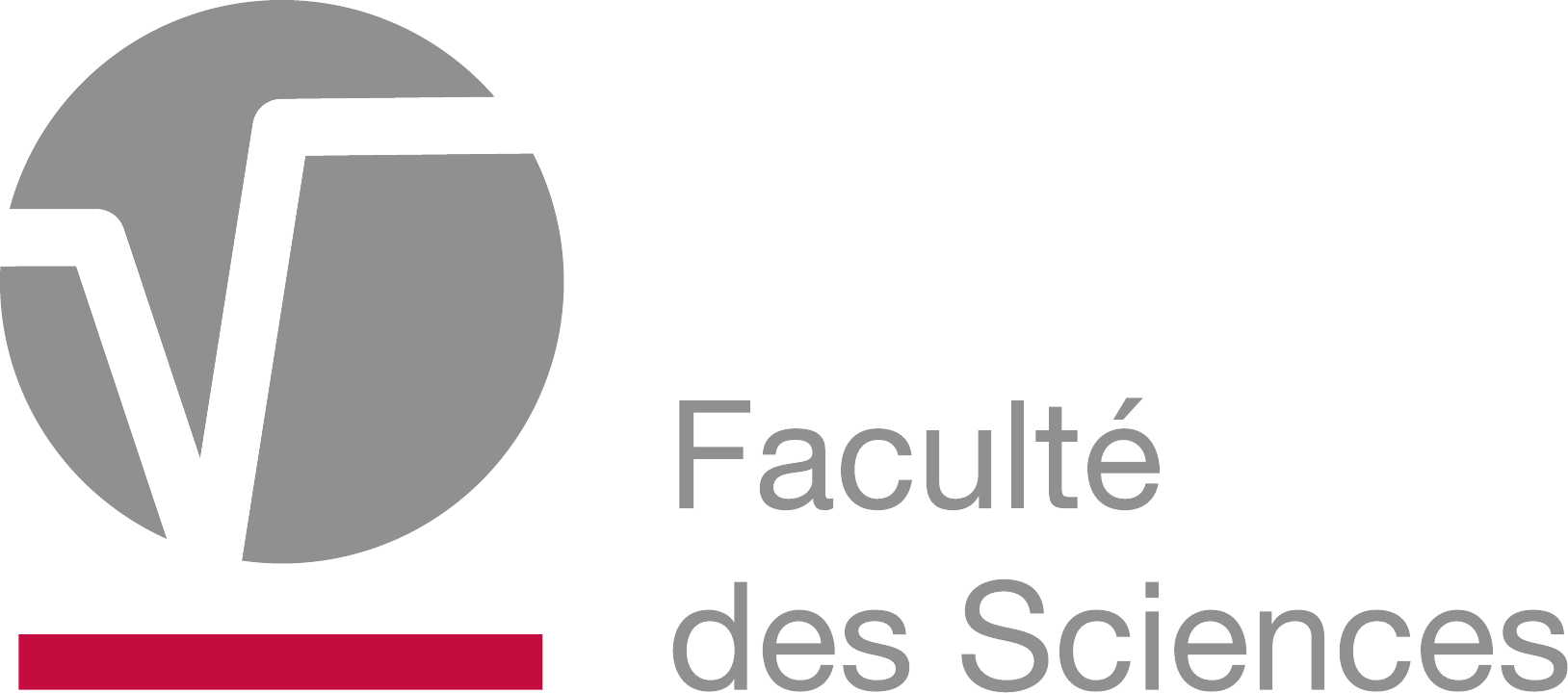}
\end{minipage}
\end{figure}
\vspace*{1cm}
{\scshape\LARGE \univname\par}\vspace{1cm} 
\textsc{\Large Master Thesis}\\[0.5cm] 
\vspace*{0.5cm}
\HRule \\[0.4cm] 
{\fontsize{0.55475cm}{0.7cm}\selectfont \bfseries \ttitle\par}\vspace{0.4cm} 
\HRule \\[1.5cm] 
 
\begin{minipage}[t]{0.4\textwidth}
\begin{flushleft} \large
\emph{Author:}\\
\authorname
\\\small\href{mailto:\authmail}{\authmail} 
\end{flushleft}
\end{minipage}
\begin{minipage}[t]{0.4\textwidth}
\begin{flushright} \large
\emph{Supervisor:} \\
\supname
\\\small\href{mailto:\supmail}{\supmail} 
\end{flushright}
\end{minipage}\\[2cm]
 
\vfill

\large \textit{A thesis submitted in fulfillment of the requirements\\ for the degree of \degreename}\\[0.3cm] 
\textit{in the}\\[0.4cm]
\groupname\\\deptname\\[1.2cm] 
 
\vfill

{\large Academic year 2016-2017}\\[.5cm] 
 
\vfill
\smash{\raisebox{-5mm}{\vtop{%
			\small
			\facname\bottomsep
			\univname\bottomsep
			Place du Parc 20\bottomsep
			B-7000 Mons 
}}}%
\end{center}
\end{titlepage}

\begin{abstract}
\addchaptertocentry{\abstractname} 
Achieving a symbiotic blending between reality and virtuality is a dream that has been lying in the minds of many people for a long time. Advances in various domains constantly bring us closer to making that dream come true. Augmented reality as well as virtual reality are in fact trending terms and are expected to further progress in the years to come. 

This master's thesis aims to explore these areas and starts by defining necessary terms such as augmented reality (AR) or virtual reality (VR). Usual taxonomies to classify and compare the corresponding experiences are then discussed.

In order to enable those applications, many technical challenges need to be tackled, such as accurate motion tracking with 6 degrees of freedom (positional and rotational), that is necessary for compelling experiences and to prevent user sickness. Additionally, augmented reality experiences typically rely on image processing to position the superimposed content. To do so, “paper” markers or features extracted from the environment are often employed. Both sets of techniques are explored and common solutions and algorithms are presented.

After investigating those technical aspects, I carry out an objective comparison of the existing state-of-the-art and state-of-the-practice in those domains, and I discuss present and potential applications in these areas. As a practical validation, I present the results of an application that I have developed using Microsoft HoloLens, one of the more advanced affordable technologies for augmented reality that is available today. Based on the experience and lessons learned during this development, I discuss the limitations of current technologies and present some avenues of future research.
\end{abstract}


\blankpage
\addchaptertocentry{\acknowledgementname} 
\phantom{.}
\vfill
\begin{margin}{5cm}{0cm} 
	\qquad I would like to thank:
	
	\qquad Prof. Tom Mens for allowing me to explore a field of research I was particularly interested in, as well as lending me the needed devices and providing advice,
	
	\qquad The Microsoft Innovation Center Belgium and in particular Frédéric Carbonnelle for the opportunity of using the Hololens, even letting me borrow such an innovative device during an entire weekend for the "Printemps des Sciences",
	
	\qquad My colleagues from the entrepreneurial project for exploring the business aspect of augmented reality,
	
	\qquad And finally, my family for supporting me all along.
\end{margin}
\blankpage


\setcounter{tocdepth}{2}
\tableofcontents 

\listoffigures 

\listoftables 


\begin{abbreviations}{lll} 

\textbf{Abbreviation} 		& \textbf{Meaning} 		& \textbf{Defined in:} \\
\hline\\
\textbf{AEC}		& \textbf{A}rchitecture \textbf{E}ngineering and \textbf{C}onstruction 							& section \ref{vr-CAD}\\							
\textbf{AR} 		& \textbf{A}ugmented \textbf{R}eality  		        											& section \ref{ar-def}\\		
\textbf{ASIC}		& \textbf{A}pplication-\textbf{S}pecific \textbf{I}ntegrated \textbf{C}ircuits 					& section \ref{limitations-computing}\\							
\textbf{AV} 		& \textbf{A}ugmented \textbf{V}irtuality  														& section \ref{av-def}\\	
\textbf{BCI} 		& \textbf{B}rain-\textbf{C}omputer \textbf{I}nterfaces     										& section \ref{limitations-interaction}\\					
\textbf{BRIEF}		& \textbf{B}inary \textbf{R}obust \textbf{I}ndependent \textbf{E}lementary \textbf{F}eatures  	& section \ref{natural-feature}\\							
\textbf{BRISK}		& \textbf{B}inary \textbf{R}obust \textbf{I}nvariant \textbf{S}calable \textbf{K}eypoints  		& section \ref{natural-feature}\\							
\textbf{CAAD} 		& \textbf{C}omputer-\textbf{A}ided \textbf{A}rchitectural \textbf{D}esign 						& section \ref{vr-CAD}\\					
\textbf{CAD} 		& \textbf{C}omputer-\textbf{A}ided \textbf{D}esign       										& section \ref{vr-CAD}\\				
\textbf{CAVE} 		& \textbf{C}ave \textbf{A}utomatic \textbf{V}irtual \textbf{E}nvironment						& section \ref{vr-def}\\					
\textbf{DoF} 		& \textbf{D}egrees \textbf{o}f \textbf{F}reedom 												& section \ref{motion-tracking}\\			
\textbf{DoG} 		& \textbf{D}ifference \textbf{o}f \textbf{G}aussians 											& section \ref{natural-feature}\\			
\textbf{DSP} 		& \textbf{D}igital \textbf{S}ignal \textbf{P}rocessor       									& section \ref{limitations-computing}\\					
\textbf{EEG} 		& \textbf{E}lectro\underline{\smash{e}}ncephalo\underline{\smash{g}}rams       					& section \ref{limitations-interaction}\\					
\textbf{EKF} 		& \textbf{E}xtended \textbf{K}alman \textbf{F}ilter       										& section \ref{slam}\\				
\textbf{EMG} 		& \textbf{E}lectro\underline{\smash{m}}yo\underline{\smash{g}}rams     							& section \ref{limitations-interaction}\\					
\textbf{EPM} 		& \textbf{E}xtent of \textbf{P}resence \textbf{M}etaphor 										& section \ref{other-taxonomies}\\				
\textbf{EWK} 		& \textbf{E}xtent of \textbf{W}orld \textbf{K}nowledge     										& section \ref{other-taxonomies}\\
\textbf{FAST}		& \textbf{F}eatures from \textbf{A}ccelerated \textbf{S}egment \textbf{T}est 					& section \ref{natural-feature}\\							
\textbf{FLANN}		& \textbf{F}ast \textbf{L}ibrary for \textbf{A}pproximate \textbf{N}earest \textbf{N}eighbors  	& section \ref{natural-feature}\\							
\textbf{FoV}		& \textbf{F}eld \textbf{o}f \textbf{V}iew  														& section \ref{ar-def}\\							
\textbf{FREAK}   	& \textbf{F}ast \textbf{Re}tin\textbf{a} \textbf{K}eypoints        								& section \ref{natural-feature}\\							
\textbf{HMD} 		& \textbf{H}ead \textbf{M}ounted \textbf{D}isplay 												& section \ref{ar-def}\\			
\textbf{HPU} 		& \textbf{H}olographic \textbf{P}rocessing \textbf{U}nit    									& section \ref{hololens-hardware-section}\\						
\textbf{IMU} 		& \textbf{I}nertial \textbf{M}easurement \textbf{U}nit											& section \ref{inertial}\\	
\textbf{LoG} 		& \textbf{L}aplacian \textbf{o}f \textbf{G}aussian 		    									& section \ref{natural-feature}\\			
\textbf{MR} 		& \textbf{M}ixed \textbf{R}eality 		       													& section \ref{rv-continuum-section}\\
\textbf{ORB}        & \textbf{O}riented Fast \textbf{R}otated \textbf{B}rief    									& section \ref{natural-feature}\\				
\textbf{PTSD} 		& \textbf{P}ost-\textbf{T}raumatic \textbf{S}tress \textbf{D}isorder    						& section \ref{vr-healthcare}\\			
\textbf{RF} 		& \textbf{R}eproduction \textbf{F}idelity 	      												& section \ref{other-taxonomies}\\		
\textbf{RGB} 		& \textbf{R}ed \textbf{G}reen \textbf{B}lue       												& section \ref{issues}\\		
\textbf{RPD} 		& \textbf{R}etinal \textbf{P}rojection \textbf{D}isplay											& section \ref{ar-def}\\					
\textbf{RSD} 		& \textbf{R}etinal \textbf{S}canning \textbf{D}isplay											& section \ref{ar-def}\\				
\textbf{RTT} 		& \textbf{R}ound-\textbf{T}rip \textbf{T}ime       												& section \ref{issues}\\			
\textbf{RV} 		& \textbf{R}eality-\textbf{V}irtuality (continuum)	       										& section \ref{rv-continuum-section}\\	
\textbf{SAR} 		& \textbf{S}patial \textbf{A}\textbf{R} 														& section \ref{ar-def}\\		
\textbf{SIFT} 		& \textbf{S}cale-\textbf{I}nvariant \textbf{F}eature \textbf{T}ransform  						& section \ref{natural-feature}\\				
\textbf{SLAM} 		& \textbf{S}imultaneous \textbf{L}ocalization \textbf{A}nd \textbf{M}apping   					& section \ref{slam}\\			
\textbf{SMB} 		& \textbf{S}mall and \textbf{M}edium size \textbf{B}usiness         							& section \ref{ar-retail}\\					
\textbf{SURF} 		& \textbf{S}peeded \textbf{U}p \textbf{R}obust \textbf{F}eatures  								& section \ref{natural-feature}\\			
\textbf{ToF} 		& \textbf{T}ime \textbf{o}f \textbf{F}light  													& section \ref{acoustic}\\		
\textbf{UI} 		& \textbf{U}ser \textbf{I}nterface                 												& section \ref{limitations-interaction}\\		
\textbf{UWP} 		& \textbf{U}niversal \textbf{W}indows \textbf{P}latform 										& section \ref{issues}\\				
\textbf{VPU} 		& \textbf{V}ision \textbf{P}rocessing \textbf{U}nit         									& section \ref{limitations-computing}\\					
\textbf{VR} 		& \textbf{V}irtual \textbf{R}eality  		        											& section \ref{vr-def}\\		
\textbf{VRD} 		& \textbf{V}irtual \textbf{R}etinal \textbf{D}isplay 											& section \ref{ar-def}\\				
\textbf{VRET} 		& \textbf{VR} \textbf{E}xposure \textbf{T}herapy   												& section \ref{vr-healthcare}\\					

\end{abbreviations}


%
%
%


%
%
%
%




\mainmatter 

\pagestyle{thesis} 


\chapter{Introduction}

\label{intro}

Merging real and virtual worlds has been in many people's minds for decades but, as the hardware evolved (in terms of computing power and display quality) and because of advances in computer vision, such experiences are more believable every day.

While science fiction cinematographic works have helped us envision what the future could be, with futuristic virtual interfaces in movies such as Minority Report (2002) and Iron Man (2008), the aforementioned improvements make them realistic.

The huge success of Pokémon Go\footnote{\url{http://www.pokemongo.com}} as well as financial forecasts \cite{digicapital, creditsuisse} make us believe in the potential of augmented and virtual reality in many domains of human endeavor. The exact meaning of those terms will be explained in chapter \ref{def-taxonomy} and potential applications will be discussed in later chapters.

This master's thesis will first define, classify and compare categories of reality/virtuality experiences. Then, chapter \ref{tracking} will discuss several of those with regards to the techniques and technologies that are required to enable the corresponding experiences. Each classification will be illustrated with concrete applications (in chapters \ref{vr} and \ref{ar}). Chapter \ref{mr} will then present a project created as part of this master's thesis and that runs on Microsoft Hololens.

Finally, based on the experience acquired and the state of the art, chapter \ref{limitations-prospects} will dicuss current limitations as well as future prospects and research in those areas.

\chapter{Definitions and taxonomy}

\label{def-taxonomy}
Lots of papers, studies, blog articles and other resources can be found about different ways of merging reality and virtuality. Most people have heard about several terms such as augmented reality and virtual reality but what do they really mean? How do they compare/differ? This first chapter will define those terms and discuss taxonomies to understand how they relate.

\section{Augmented Reality (AR)} \label{ar-def}
\begin{figure}[h]
	\includegraphics[width=6cm]{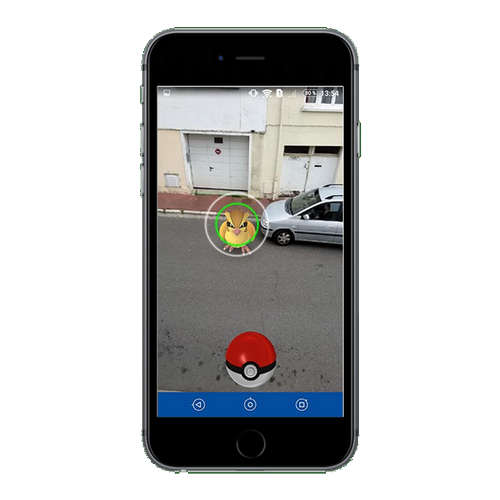}
	\centering
	\caption[AR feature in Pokémon Go]{\label{pokemongo} AR feature in Pokémon Go\protect\footnotemark }
\end{figure}

Augmented Reality (which will from now on be referred to as AR) describes the blending of the real world with a virtual one. It can basically be seen as "adding virtual things on top of the real world's perception". A typical example of an AR device is Google Glass \cite{glass} whereas the game Pokémon Go \cite{pokemongo} helped popularize AR with its integration of Pokémon on top of the live camera feed as if they were there in the real world (as shown on figure \ref{pokemongo}).

\footnotetext{\url{http://img.phonandroid.com/2016/07/pokemon-go-capture.jpg}} 

\label{displays}
In order to enable AR experiences, different kinds of displays can be used. The following sections will introduce a few of those categories and provide concrete examples.
\subsubsection{Monitor-based AR displays}
\begin{figure}
	\centering
	\begin{subfigure}[b]{7cm}            
		\frame{\includegraphics[width=6.7cm]{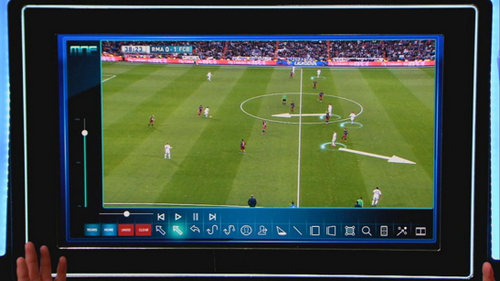}}
		\caption{Sky Sports's AR tool SkyPad as used in Monday Night Football analyses\protect\footnotemark}
		\label{ar-football}
	\end{subfigure}
	\hspace{1cm}
	\begin{subfigure}[b]{6cm}
		\centering
		\frame{\includegraphics[width=6cm]{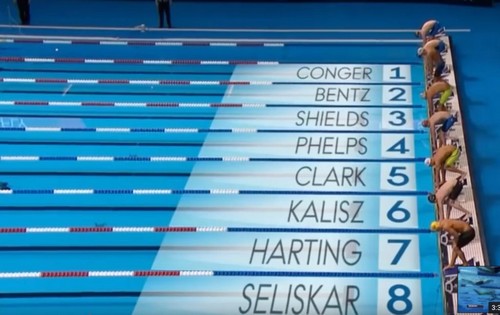}}
		\caption{Commonly seen AR feature in swimming competitions\protect\footnotemark}
		\label{ar-swimming}
	\end{subfigure}
	\caption{Examples of AR features in sports broadcasting}\label{ar-sports}
\end{figure}

This is the simplest type of AR displays available. It refers to non-immersive experiences where the augmentation happens on a "distant" screen, such as a TV or a smartphone (even though in that particular case, the more specific "handheld AR" is often used). The display is treated as a window to the augmented world (hence the alternative name "Window-on-the-World" given by \citeauthor{milgram-continuum} \cite{milgram-continuum}) created from live or stored video images. Due to its accessibility (from the user's point of view) it is the most prominent form of AR experiences, with lots of applications in sports broadcasting (examples shown in figures \ref{ar-football} and \ref{ar-swimming}). The previously mentioned feature in Pokémon Go also is an example of monitor-based AR as it takes place on smartphone screens.

\addtocounter{footnote}{-1}
\footnotetext{\url{http://e1.365dm.com/15/11/16-9/20/carra-on-real-grab_3380670.jpg?20151124084802}}
\stepcounter{footnote}
\footnotetext{\url{http://arvr-fa.ir/wp-content/uploads/2017/01/Swimming_AR-1024x646.jpeg}}

\subsubsection{See-through AR displays}
\begin{figure}[h]
	\includegraphics[width=10cm]{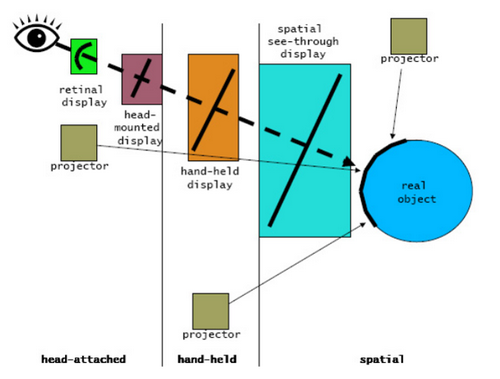}
	\centering
	\caption[Different ways of generating AR images]{\label{ar-displays} Different ways of generating AR images \cite{ar-displays}}
\end{figure}

A second class of displays used for AR experiences allows the user to see the augmented world "directly" in the sense that he sees the real world from his own perspective (as opposed to monitor-based displays that show images from a "distant" camera).
Figure \ref{ar-displays} pictures some possibilities to make the augmentation happen and highlight differences in where the observer is located in relation to the real object as well as what type of image is generated (i.e. planar or curved).

\begin{figure}[h]
	\includegraphics[width=11cm]{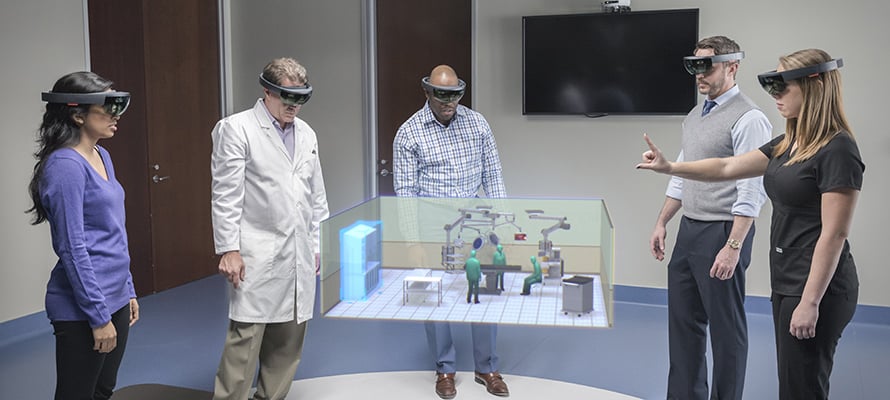}
	\centering
	\caption[Microsoft Hololens use case: Designing operating rooms with holograms integrated into the real world]{\label{hololens-example} Microsoft Hololens use case: Designing operating rooms with holograms integrated into the real world\protect\footnotemark}
\end{figure}

Head-mounted displays (HMDs) are often linked to Virtual Reality but are also extensively used in AR. In the context of a video see-through HMD, the user sees the real world through a camera system attached to the device that aims to reproduce the effective viewpoint of the user's eyes. 
It is also possible to project virtual elements on a transparent surface in front of the observer. In that case, the display is referred to as optical see-through. Depending on where that surface is located in relation to the user, different names are given: spatial see-through display if it is separated from the user's head; or simply optical see-through HMD if it is attached to it. A current example of a device falling into the latter category is Microsoft Hololens. Even though that device will be further discussed in chapter \ref{mr}, figure \ref{hololens-example} shows the idea behind it: anchoring holograms into the real world.

\footnotetext{\url{https://compass-ssl.surface.com/assets/db/25/db253de9-73e2-4a2b-a209-7a3f01972f68.jpg?n=HoloLens_Homepage_Mosaic-Stryker_1920_890x400.jpg}}

It is worth mentioning that different issues arise depending on the type of see-through being used, as pointed out by \citeauthor{fuchs-ar-displays} \cite{fuchs-ar-displays}. An optical see-through HMD will provide an unmatched direct view of the real world but will require more advanced technologies (e.g. to accommodate changes in head position/orientation rapidly enough, to deal with light intensity and to properly handle occlusion). On the other hand, a video see-through HMD will have to make sure the field of view (FOV, the extent of the scene that can be seen, generally measured in degrees) is acceptable and, more importantly, that the video itself has a high and stable framerate. Indeed, whereas a video see-through display can make sure the real world and its augmentations are synchronized, it might introduce a delay between the movements of the user's head and what he sees. This has to be handled with caution as, otherwise, users might suffer from nausea.
Another element worth pointing out is the perhaps surprising "age" of see-through displays. In fact, it appears that HMD systems were developed as early as in the late 1960s, by \citeauthor{first-ar-hmd} \cite{first-ar-hmd} and, at the same time, by \citeauthor{first-ar-hmd-army} \cite{first-ar-hmd-army} for the US Air Force with projects that then lead to the "Super Cockpit" program \cite{supercockpit, supercockpit-britannica} in \citeyear{supercockpit}.

The last kinds of see-through displays that will be discussed here are retinal displays (sometimes named RPDs for Retinal Projection Displays, RSDs for Retinal Scanning Displays or even VRDs for Virtual Retinal Displays).
As shown on the top-left corner of figure \ref{ar-displays}, it is possible to generate AR images closer to the eye. In fact, retinal displays are drawing those images directly onto the eye's retina using low-power laser beams.
While the principle could frighten readers at first glance, it should be mentioned that the eyes of the wearer remain safe, even after being exposed for several hours \cite{retinal-lew}. Even better, the technology doesn't tire the eyes as much as a conventional HMD \cite{retinal-tech-overview}.
Several prototypes (e.g. by \citeauthor{retinal-tech-binocular} from the University of Washington \cite{retinal-tech-binocular}) and even actual products (e.g. Laser EyeWear (LEW) \cite{retinal-lew} from Fujitsu, QD Laser and the University of Tokyo) exist but are not available as of yet. Despite the fact that it is not directly related to AR, it is interesting to note that such a technology can also be used to help people with low vision that "standard" glasses cannot correct \cite{retinal-low-vision}.

\subsubsection{Spatial AR}
Sometimes referred to as projection or projective AR, spatial AR (SAR) is about augmenting reality by projecting images directly onto real objects. Although figure \ref{ar-displays} already mentioned "spatial" and "projector", SAR did not really belong in the see-through category as there is no actual display other than the real world objects themselves.
First introduced by \citeauthor{sar-first} \cite{sar-first}, SAR can therefore "naturally" provide multi-user experiences. As of now, most applications of those techniques are related to a cultural context where images are projected onto surfaces such as the facade of a building. Sometimes referred to as monumental projections \cite{monumental-projection} and video or 3D mapping, those techniques can also provide 360\textdegree experiences \cite{sar-mons2015} as well as user's interaction \cite{sar-polytech}.

\subsubsection{Other types of AR}
So far only visual AR categories have been presented. It should however be mentioned that other kinds of AR exist that are using different senses: audio, haptic, olfactory and gustatory AR. Haptic (touch) interfaces will briefly be discussed in chapter \ref{limitations-prospects} but audio, olfactory and gustatory AR go beyond the scope of this work.

\section{Augmented Virtuality (AV)} \label{av-def}
\begin{figure}[h]
	\includegraphics[width=8cm]{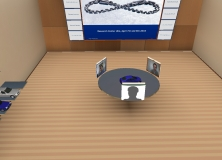}
	\centering
	\caption[cAR/PE!: AV videoconferencing system]{\label{av-conference} cAR/PE!: AV videoconferencing system by \citeauthor{av-conference} \cite{av-conference}}
\end{figure}

While AR was about augmenting the real world with virtual objects, augmented virtuality (AV) is about adding real world elements into a virtual world.
Most AV applications can be subsumed as "chroma key experiences". Examples include weather forecast broadcasts and video conferences into virtual environments \cite{av-conference} (as shown in figure \ref{av-conference}).
Even closer to virtual reality (discussed in section \ref{vr-def}), applications where the user's hands are integrated into a virtual world (e.g. a "virtual studio for architectural exploration" by \citeauthor{av-hands} \cite{av-hands}) are also examples of AV.

As of now, AV is nowhere near as popular as AR or VR. In fact, it is highly unlikely that this situation will change in the future for two reasons. Firstly, because the boundaries between those categories can be "blurred" and people will naturally use terms they know. Secondly, because technologies will keep on improving and it will become harder and harder to distinguish what is physically real from what was virtually added.
\section{Virtual Reality (VR)} \label{vr-def}
\begin{figure}[h]
	\includegraphics[width=8cm]{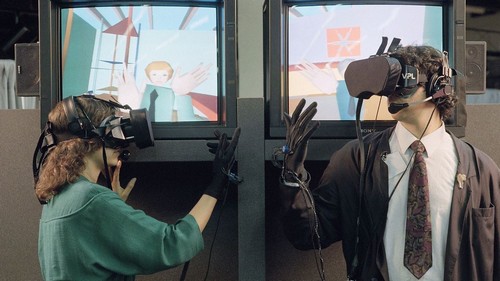}
	\centering
	\caption[The EyePhone: one of the very first VR HMDs available for purchase]{\label{vr-first-hmd} The EyePhone: one of the very first VR HMDs available for purchase\protect\footnotemark}
\end{figure}

\begin{figure}[h]
	\includegraphics[width=8cm]{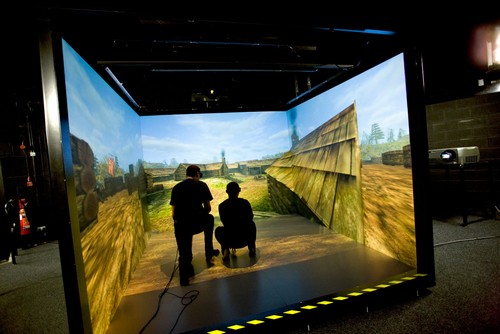}
	\centering
	\caption[A recent setup of a CAVE-like setup at Teesside University]{\label{vr-cave} A recent setup of a CAVE-like setup at Teesside University\protect\footnotemark}
\end{figure}

Virtual Reality (VR) describes experiences where the user is entirely immersed into a three-dimensional virtual world and interacts with it.
While the paternity of the concept is unclear, VR as a term is generally attributed to \citeauthor{vr-origins-itw} who worked actively \cite{vr-origins, vr-origins-itw} in the domain in the late 1980's.
The usual equipment used for VR experiences involves a HMD and some kind of controller (in early products from \citeauthor{vr-origins-itw}: a glove). Figure \ref{vr-first-hmd} shows one of the first VR commercial products by VPL Research (\citeauthor{vr-origins-itw}'s company) in 1989.
\footnotetext{\url{http://i.amz.mshcdn.com/auQDd-I-BKnVdiXWK7r53vtDA_g=/fit-in/1200x9600/2014\%2F04\%2F14\%2F03\%2FEyephoneVPL.99ebb.jpg}}
\newpage
It should however be mentioned that VR is not limited to HMDs. In the early 1990's, \citeauthor{vr-cave} at the University of Illinois developed CAVE (Cave Automatic Virtual Environment) \cite{vr-cave}, a VR setup in a cubic room, with images projected on the walls. This alternative to HMDs provides great immersive experiences with a wide field-of-view (see figure \ref{vr-cave}) but requires extra room and isn't as affordable.

\footnotetext{\url{http://static.worldviz.com/_wp/wp-content/uploads/2012/09/CAVE_christie_2.jpg}}

\section{The Reality-Virtuality continuum} \label{rv-continuum-section}
\begin{figure}[h]
	\includegraphics[width=13cm]{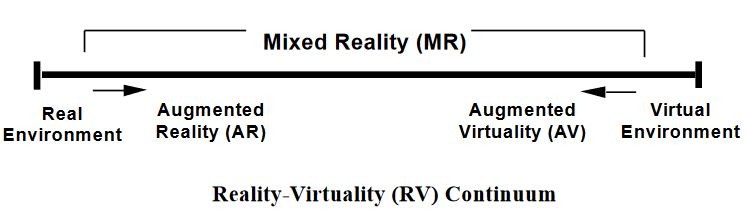}
	\centering
	\caption[The original RV continuum by \citeauthor{rv-continuum}]{\label{rv-continuum} The original RV continuum by \citeauthor{rv-continuum} \cite{rv-continuum}}
\end{figure}

Now that some terms have been defined, it is time to introduce the most frequently used taxonomy: the Reality Virtuality continuum by \citeauthor{rv-continuum} \cite{rv-continuum}.
In 1994, AR was already a popular term in the literature but different definitions where given. Therefore, \citeauthor{rv-continuum} wanted to clarify what that term meant and how it related to VR. 
In order to do so, they created a continuum (shown in figure \ref{rv-continuum}) that is still considered as the main reference to classify experiences mixing real and virtual elements.

The idea behind this one-dimensional taxonomy is that there is a broad range of applications between an entirely real world and a solely virtual environment. They can be placed on that axis depending on whether they are primarily using reality or the virtual world.

In that figure, the term Mixed Reality (MR) is introduced but we (purposely) did not define it yet. In fact, it has been misused in recent years to describe "spatial aware" AR devices and experiences (further discussion on that in chapter \ref{mr}). \citeauthor{rv-continuum} defines a MR environment as \enquote{one in which real world and virtual world objects are presented together within a single display, that is, anywhere between the extrema of the RV continuum}. Therefore, MR is a subset containing AR, AV and even VR experiences as they are not entirely virtual because section \ref{vr-def} stated that, in order to be considered as a VR experience, an application needs to include user interactions.

\begin{figure}[h]
	\includegraphics[width=13cm]{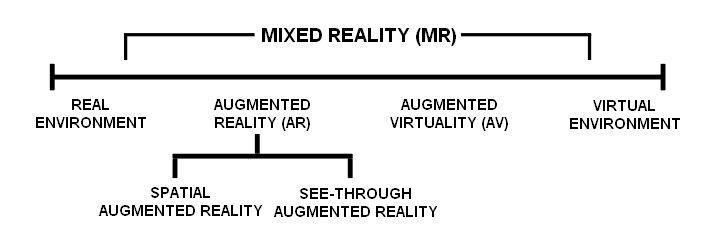}
	\centering
	\caption[The extended RV continuum by \citeauthor{sar-thesis}]{\label{rv-continuum-sar} The extended RV continuum by \citeauthor{sar-thesis} \cite{sar-thesis}}
\end{figure}

In his PhD thesis about spatial AR, \citeauthor{sar-thesis} \cite{sar-thesis} proposed an extended version of \citeauthor{rv-continuum}'s RV continuum to differentiate SAR from see-through AR. As SAR directly projects the images on a real world surface, he chose to put it further the left. Figure \ref{rv-continuum-sar} shows that version of the continuum.

\begin{figure}[h]
	\includegraphics[width=14cm]{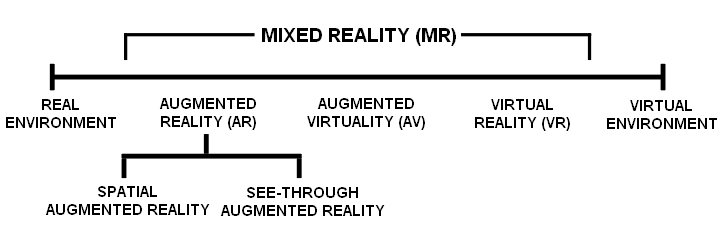}
	\centering
	\caption[Our proposal: A further extended RV continuum including VR]{\label{rv-continuum-proposal} Our proposal: A further extended RV continuum including VR}
\end{figure}

In order to clarify where VR experiences should be placed on that continuum, we slightly modified \citeauthor{sar-thesis}'s version. The resulting further extended RV continuum is shown in figure \ref{rv-continuum-proposal}.

Even though some terms mentioned previously are not included in it (e.g. we could also differentiate between different kinds of see-through AR), we believe this version is sufficiently complete to clarify everything that has been mentioned so far.

\section{Other taxonomies} \label{other-taxonomies}
\begin{figure}[h]
	\includegraphics[width=6cm]{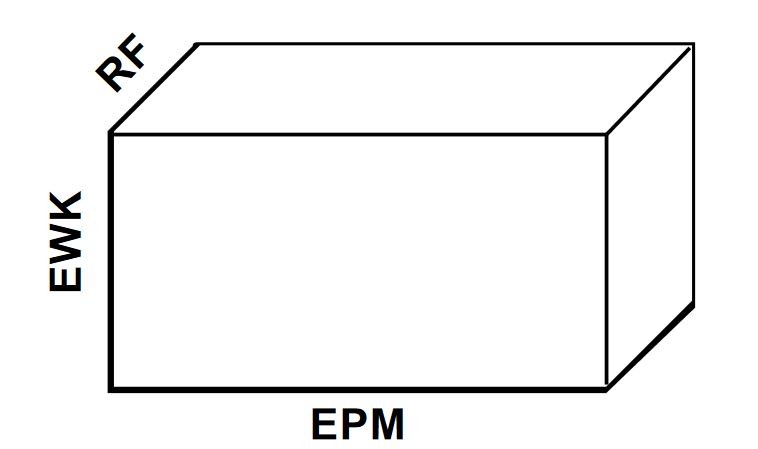}
	\centering
	\caption[\citeauthor{milgram-continuum}'s three-dimensional taxonomy for classifying mixed reality displays]{\label{three-dim-mr-taxonomy} \citeauthor{milgram-continuum}'s three-dimensional taxonomy for classifying mixed reality displays \cite{milgram-continuum}}
\end{figure}

As a one-dimensional continuum is not sufficient to highlight the differences between a plethora of mixed reality experiences, other taxonomies are needed.
In fact, in the same paper \cite{milgram-continuum}, \citeauthor{milgram-continuum} summarize the main points of a three-dimensional taxonomy for mixed reality systems, that is further discussed in another paper \cite{mr-taxonomy}.
The paper describes 3 axes: Extent of World Knowledge (EWK: how much do we know about the - real or virtual - world in which the experience happens?), Reproduction Fidelity (RF: is the augmented content realistic?) and Extent of Presence Metaphor (EPM: is the user immersed in the experience or does he look at a monitor?). The resulting three-dimensional "hyperspace" is depicted in figure \ref{three-dim-mr-taxonomy}.

Another classification worth mentioning is Fuchs' VR taxonomy \cite{traite-vr-vol1} based on what he describes as its inherent functions: the ability for the user to pull himself out of his own real environment $(1)$ and/or his present time $(2)$ in order to interact $(3)$ in a virtual world.
\\
\\
\begin{minipage}{\textwidth}
	
	$(1)$ leads to states where the user is:
	\\
	\\
	$\begin{cases}
	L_\rightarrow & \text{in a distant or non-human scale environment}\\
	L_\cup & \text{in a virtual world with other users}\\
	L_0 & \text{in the same place or the location is irrelevant to the application}\\
	\end{cases}$
	
\end{minipage}
\\
\\

\begin{minipage}{\textwidth}
	
$(2) \text{ leads to states where the user is:}\begin{cases}
T_0 & \text{in present time}\\
T_- & \text{in the past}\\
T_+ & \text{in the future}\\
\end{cases}$

\end{minipage}
\\
\\

\begin{minipage}{\textwidth}
	
$(3) \text{ leads to states where the user is:}\begin{cases}
\mathit{IA}_r & \text{in a world that simulates reality}\\
\mathit{IA}_i & \text{in an imaginary/symbolic world}\\
\mathit{IA}_0 & \text{in the real world}\\
\end{cases}$

\end{minipage}
\\
\\

Experiences can then be classified by combining those characteristics (14 resulting categories because multiple users ($L_\cup$) can only interact in the present moment ($L_0$)). Virtual tours are therefore in $(\mathit{IA}_r,T_0,L_\rightarrow)$ and a VR sci-fi game would be in $(\mathit{IA}_i, T_+, (L_0 + L_\rightarrow))$, with the $+$ sign meaning "and/or".

Many other taxonomies have been proposed. A few more examples are classifications based on what is being augmented \cite{taxonomy-mackay, taxonomy-hughes} (user, objects or environment), by purpose \cite{taxonomy-dubois}, or based on whether the action is determined by the system or the user \cite{taxonomy-renevier}. However, \citeauthor{milgram-continuum}'s RV continuum remains the standard reference in the domain and will therefore be used throughout this thesis.
\chapter{Motion tracking and related computer vision techniques}

\label{tracking}

VR and AR have shared requirements in terms of motion and positional tracking, common techniques to meet those needs will therefore be presented. Additionally, AR applications often have to recognize landmarks in the environment and get a sense of the user's surroundings, the corresponding problems and specific computer vision techniques to tackle them will also be discussed in this chapter.

\section{Motion tracking} \label{motion-tracking}

In order to enable compelling VR applications, a way to track the user's head (and optionally some kind of controller) is needed. This section will present a few techniques that can be used for that matter but first, let us clarify what we need exactly.

\begin{figure}[h]
	\includegraphics[width=9cm]{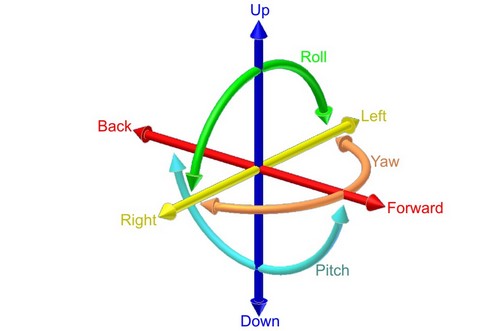}
	\centering
	\caption[6 degrees of freedom (DoF)]{\label{6dof} 6 degrees of freedom (DoF)\protect\footnotemark }
\end{figure}

We want to be able to track movements in 3D space, which means we need to know the tracked object's position as well as its rotation. That kind of tracking is usually labeled as 6 DoF (degrees of freedom) because the object effectively has 6 separate ways to modify its position (3 axes: x, y, z) and rotations (yaw, pitch, roll) as seen in figure \ref{6dof}, where each color represents one of those degrees of freedom.

\footnotetext{\url{https://upload.wikimedia.org/wikipedia/commons/f/fa/6DOF_en.jpg}} 

An ideal tracking system would need to meet the following characteristics:
\begin{itemize}
	\item High accuracy: user motions should be accurately tracked
	\item High precision: jitter should be negligible and imperceptible by the user, no movement should be registered when the tracking target is not moving
	\item Low latency and high update rate: the delay between a motion and the system's awareness of that motion should be very low as failing to do so leads to user sickness
	\item Wide range: users should not be restricted to a small area of interaction around the sensor(s)
	\item High mobility: users should be able to move freely (wireless and autonomous trackers are therefore preferred) in their environment which also implies the tracking system has to be as small and light as possible
	\item Environmental robustness: disruptive elements from the environment (e.g. sunlight, temperature or magnetic fields) should not alter the tracking system's quality
\end{itemize}

\subsection{Mechanical tracking}
\begin{figure}[h]
	\includegraphics[width=10cm]{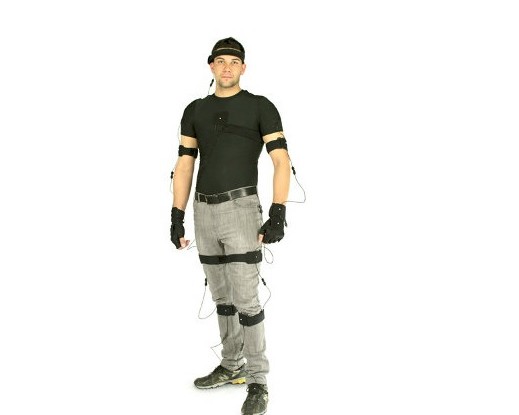}
	\centering
	\caption[MotionShadow, a full body tracking system]{\label{body-tracking} MotionShadow, a full body tracking system\protect\footnotemark }
\end{figure}

Mechanical tracking probably is simpler than other methods, at least conceptually. In fact, the tracked object is typically directly linked to the system via several mechanical "arms" (made up of articulated pieces). The object's position and rotation is then determined using the angles of the arms' joints (with sensors placed on those joints). The same principle (calculating the angles of specific joints) has also been used for full body tracking with complete suits (that are very expensive but typically wireless) such as Inition's MotionShadow\footnote{\url{https://www.inition.co.uk/product/motionshadow-full-body-tracking-system/}} pictured in figure \ref{body-tracking}.

\footnotetext{\url{https://www.inition.co.uk/wp-content/uploads/2014/11/Shadow_full-body-519x415.jpg}} 

The issue with that solution therefore is either the cost (for body suits) or the restricted movements provided by mechanical structures the user is attached to.

\subsection{Magnetic tracking}
\begin{figure}[h]
	\includegraphics[width=5cm]{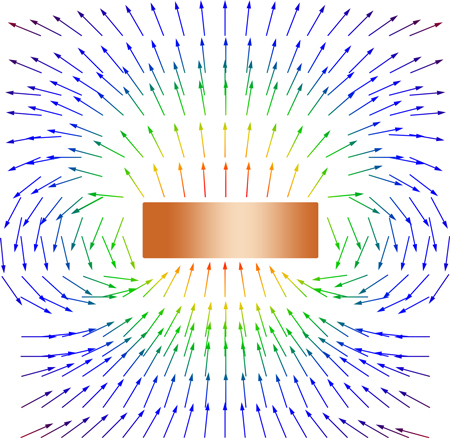}
	\centering
	\caption[Example of magnetic field lines]{\label{magnetic} Example of magnetic field lines\protect\footnotemark }
\end{figure}

Magnetic tracking uses a base station that generates current and therefore a magnetic field. Sensors are placed on the tracked device and are able to measure the magnitude of that magnetic field (it varies depending on the distance) as well as its direction (an example of a magnetic field can be seen in figure \ref{magnetic}). 

Using that data, both the position and the rotation of the device can be determined.
Example of commercial products using that technology include Razer Hydra controllers\footnote{\url{http://www.razersupport.com/gaming-controllers/razer-hydra/}} (now classified as "legacy") and Polhemus trackers\footnote{\url{http://polhemus.com/applications/electromagnetics}}.

\footnotetext{\url{http://www.allegromicro.com/~/media/Images/Design/Position-And-Level-Sensing-Using-Hall-Effect-Sensing-Technology/fig9.ashx?w=450\&h=438\&as=1\&la=en\&hash=D2BFDDB36804067CF413236AFC931D8DBBCCE12D}} 

That solution means that the base station does not need to be in sight (with regards to the sensors) but other devices (electrical appliances, computers, etc) can cause disturbances.

\subsection{Acoustic tracking} \label{acoustic}
Another possibility to track those 6 DoF uses ultrasonic ($>20000 \textit{ Hz}$) signals emitted by several sources. Receivers are placed on the tracked object and Time of Flight (ToF) measurements determine the distance from the emitter to each receiver. As the sensors' geometry on the tracked object is known, the position and rotation can be computed. Even though it isn't the most popular method in VR, a few commercial products exist such as some InterSense trackers\footnote{\url{http://www.intersense.com/}}.

That solution is cheap but isn't very resistant to environmental interference (e.g. pressure or temperature changes)
\subsection{Inertial tracking} \label{inertial}
Using accelerometers and gyroscopes (inertial sensors as they are based on the principle of inertia i.e. $F = ma = m\frac{dv}{dt}$), one can also get a 6-DoF tracking solution.
An accelerometer measures the difference between the object's acceleration projected on the sensitive axis and gravity ($9.81$ m/s² upwards), thus enabling one-dimensional position tracking, whereas gyroscopes measures rotation around a single axis.
Resulting measurements (from 3 accelerometers and 3 gyroscopes) can then be used to provide both the position and the rotation of an object, in what is generally called an IMU (inertial measurement unit).

\begin{figure}[h]
	\includegraphics[width=6cm]{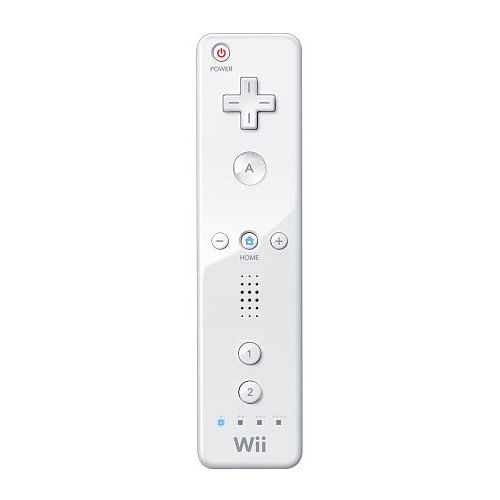}
	\centering
	\caption[Nintendo's controller, the Wiimote]{\label{wiimote} Nintendo's controller, the Wiimote\protect\footnotemark }
\end{figure}

That solution is very widely available as smartphones contain those sensors, most people therefore have a 6 DoF tracking system in their pocket (although only rotation data is relatively reliable, because of the drifting problem described below).
Other products using those principles include Nintendo's Wiimote (that also uses optical tracking, discussed in section \ref{optical-tracking}) pictured in figure \ref{wiimote} as well as Microsoft Hololens (see section \ref{hololens}).

\footnotetext{\url{https://vignette4.wikia.nocookie.net/ssb/images/0/0c/WiiMote.jpg/revision/latest?cb=20070925042013}} 

The biggest advantage of using inertial tracking is that it doesn't require any base station or emitter but one should note that it also is immune to environmental interference. However, they are either very expensive (compared to other solutions) or less accurate. They are often coupled with another method because accelerometers cannot be used for long-term tracking on their own (as demonstrated in \cite{drift-vid}). They in fact drift and because accelerometers measure acceleration, from which a relative position is derived, offset errors are accumulated quadratically.

\subsection{Optical tracking} \label{optical-tracking}
Optical solutions are probably the most diverse and widely used tracking systems. The general principle is that some kind of sensor (typically some kind of camera) will either track features/patterns in the environment or active/passive markers and use them to determine the object's position and rotation. The camera(s) can either be "external" to the tracked object (outside-in tracking) or attached to it (inside-out tracking).

\subsubsection{Active and passive markers} 
Solutions using active markers rely on light-emitting sources such as LEDs or simple light bulbs that are placed on the tracked object. Those lights are often invisible to humans (infrared lights that are only visible to infrared sensors) but need a power source. Depending on how and what one needs to track, the need for batteries can be an issue.

\begin{figure}[h]
	\includegraphics[width=8cm]{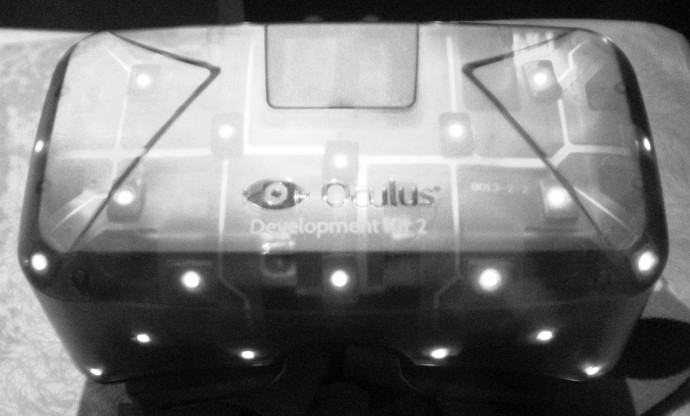}
	\centering
	\caption[Active markers on the Oculus Rift DK2]{\label{oculus-markers} Active markers on the Oculus Rift DK2\protect\footnotemark }
\end{figure}
\footnotetext{\url{http://3dvision-blog.com/wp-content/uploads/2014/09/oculus-rift-development-kit-2-ir-markers.jpg}} 

Passive markers are not luminescent themselves but often reflect light emitted by a source. The light emitted and its reflection typically are invisible to the human eye as well. "Paper" markers can be used as well and are classified as passive markers.

As previously mentioned, it is also possible to dispose of markers entirely e.g. using features from the environment or a projected pattern.

\subsubsection{Invisible light}
Using light sources typically leads to better results but it is generally desirable not to distract the user with them. Infrared lights are "too red" for the human eye to see them and are therefore often used as "invisible" lights. It should however be mentioned that using several infrared sensors can be a problem (e.g. overlapping projected patterns confusing the sensors) and that there still are drawbacks (e.g. sunlight interferences and issues with certain types of surfaces).
A few examples of techniques using infrared lights are given below.

\begin{figure}[h]
	\includegraphics[width=5cm]{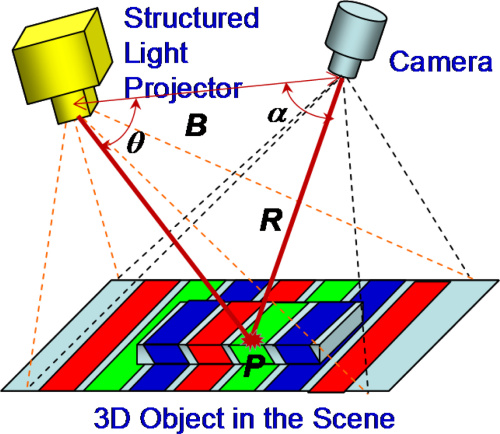}
	\centering
	\caption[The structured light principle used in Kinect v1]{\label{structured-light} The structured light principle used in Kinect v1\protect\footnotemark }
\end{figure}

The first version of Microsoft's Kinect camera uses a technique called "structured light", which essentially projects infrared patterns
to get depth information. With that data, it can compute a 3D representation of the scene it sees. Figure \ref{structured-light} shows the principle behind the method: the projected pattern will be distorted according to its distance from the source. A more detailed explanation using animations can be found in \cite{how-kinect}.

\footnotetext{\url{http://imagebank.osa.org/getImage.xqy?img=QC5sYXJnZSxhb3AtMy0yLTEyOC1nMDAx}} 

\begin{figure}[h]
	\includegraphics[width=13cm]{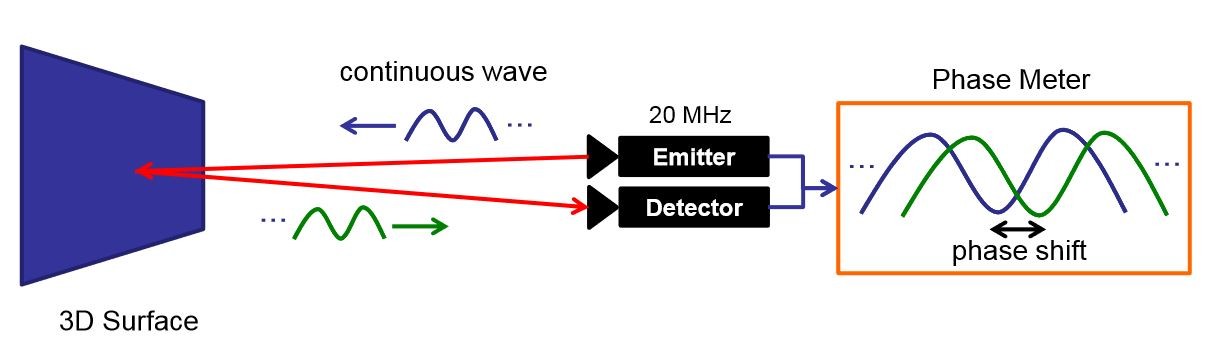}
	\centering
	\caption[The ToF continuous wave technology used in Kinect v2]{\label{tof-cw} The ToF continuous wave technology used in Kinect v2\protect\footnotemark }
\end{figure}

Released in late 2013, the second generation, “Kinect for Xbox One” also known
as Kinect v2, employs a lightly different technology to get depth data: the time of flight (ToF) measurements we talked about in section \ref{acoustic}. The light emitted by the sensor is reflected on many kinds of surfaces and the time it takes to return to the sensor is measured. As light speed is a well known and constant value, the distance from the sensor can be computed. The principle is shown in figure \ref{tof-cw}, where one can see that time is in fact measured using phase shifts.

\footnotetext{\url{https://hsto.org/getpro/habr/post_images/0e0/0ce/ff6/0e00ceff672a747577bb86fbd031242e.png}} 

Another popular example of an optical tracking system is Valve's Lighthouse with its base station(s), often paired with HTC Vive's headset and controllers. A high-level view of how the optical tracking works can be found in \cite{lighthouse-vid} and \citeauthor{lighthouse-blog}'s article \cite{lighthouse-blog} provides a lot more details as well as precision and accuracy evaluations.

Each base station uses two lasers to project lines of light (one laser horizontally; the other one vertically, with regards to the station's coordinate system). As only one laser can emit light at any time, the base stations are synchronized using flashing LEDs. The same flashing LEDs are used to synchronize tracked devices.

\begin{figure}[h]
	\includegraphics[width=10cm]{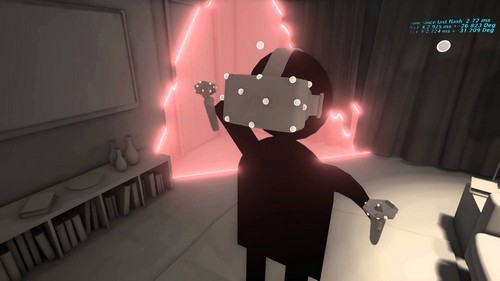}
	\centering
	\caption[Valve's Lighthouse illustrated with an HTC Vive headset and controllers]{\label{lighthouse} Valve's Lighthouse illustrated with an HTC Vive headset and controllers\protect\footnotemark }
\end{figure}

The idea (pictured in figure \ref{lighthouse}) is that when it receives a synchronization pulse, a receptor starts counting. When a laser beam hits a sensor, the counting stops and the object's position is updated (on a single axis as lines of lights are either horizontal or vertical). In between those "optical updates", an IMU is used to estimate the object's position (one of Lighthouse's purposes is therefore to correct the IMU's inevitable drifting).

\footnotetext{\url{https://i.ytimg.com/vi/J54dotTt7k0/maxresdefault.jpg}} 

\subsection{Hybrid solutions}

As \citeauthor{motion-tracking} imply in their paper's title (\citetitle{motion-tracking} \cite{motion-tracking}) and as seen in previous sections, every method has its drawbacks. For that reason, lots of trackers are combining different technologies in hybrid solutions. The term "sensor fusion" is often coined to refer to combining sensory data.
In order to blend that data, several algorithms can come to help. A very frequent choice is Kalman filtering (originally described in \citeyear{kalman} \cite{kalman}) that relies on a mathematical model to filter signals with an acceptable amount of statistical noise and inaccuracies.

One should be aware that, sometimes, hybrid trackers are described as $n$-DoF solutions, with $n > 6$. As there only exist 6 degrees of freedom for rigid bodies, this is simply a marketing trick to indicate that different sensors are tracking the same degrees of freedom. For example, IMUs are frequently equipped with magnetometers in addition to the accelerometers and gyroscopes. Those magnetometers can be used as compasses to "reset" gyroscopes' drift. As each type of sensor (3-axis accelerometer, 3-axis gyroscope and 3-axis magnetometer) measures 3 degrees of freedom, the device is occasionally labeled as a 9-DoF tracking system (when in fact gyroscopes and magnetometers are measuring the same degrees of freedom).

\section{Vision-based tracking}

The set of techniques discussed in section \ref{motion-tracking} and their combination are sufficient for VR and motion capture.
However, AR applications typically rely on image processing and often use "paper" markers (they can be used for motion tracking but other techniques are generally preferred) or features detected in the environment. While vision-based tracking is a form of optical tracking and could therefore be placed in section \ref{optical-tracking}, the variety of techniques in use and their predominance in AR deserve a specific section.

\subsection{Template markers} 

\begin{figure}[h]
	\includegraphics[width=10cm]{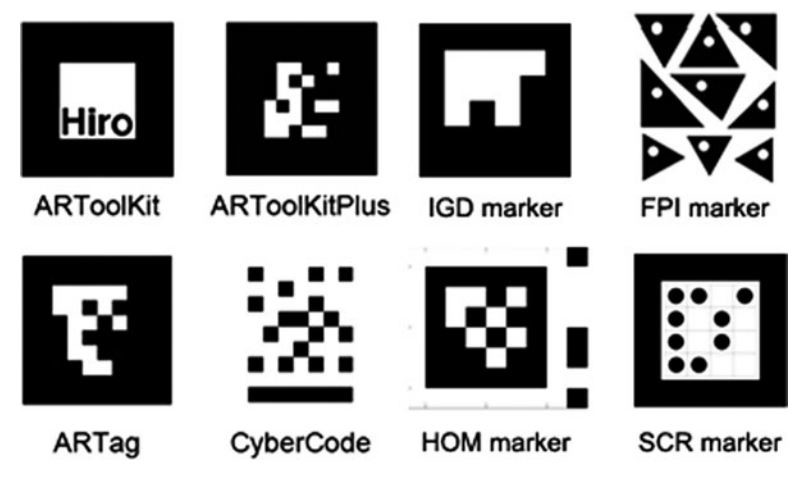}
	\centering
	\caption[Some well-known square patterns]{\label{ar-markers} Some well-known square patterns \cite{qrcode-ar}}
\end{figure}

Traditional vision-based tracking uses template markers, generally square patterns as they are believed to be the best choice to position augmented content, with regards to the criteria chosen in \cite{best-fiducial} by \citeauthor{best-fiducial}. A few examples of well-known square markers are shown in figure \ref{ar-markers}.

\begin{figure}[h]
	\includegraphics[width=13cm]{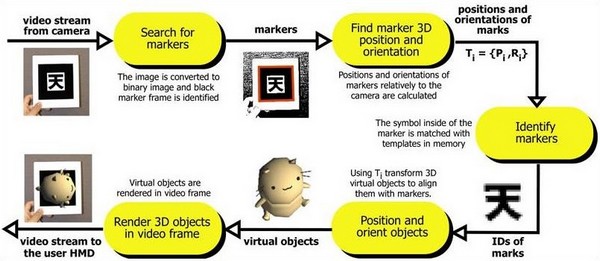}
	\centering
	\caption[ARToolkit's work-flow to superimpose augmented content on traditional template square markers]{\label{artoolkit-diagram} ARToolkit's work-flow to superimpose augmented content on traditional template square markers\protect\footnotemark }
\end{figure}

ARToolkit\footnote{\url{https://artoolkit.org}} probably is the most popular open-source AR tool. Figure \ref{artoolkit-diagram} helps explaining how the system works.

It first searches for black square shapes. Once one has been found, its inner content is analyzed. If that embedded content matches an expected pattern (the exact marker has thus been identified), the corresponding content is superimposed on it (with proper scale, position and orientation by using the marker's corners).

\footnotetext{\url{https://artoolkit.org/documentation/lib/exe/detail.php?id=3_Marker_Training\%3Amarker_about&media=diagram.jpg}} 

Different kinds of markers have been utilized with interesting properties for some use cases. For example, a QR Code can easily be identified and contains encoded data, which means resulting AR applications could read the code to get a URL pointing to a remote model to be downloaded (and then superimposed on the video feed). This approach has been discussed and tested in \cite{qrcode-ar}.

\begin{figure}[h]
	\includegraphics[width=11cm]{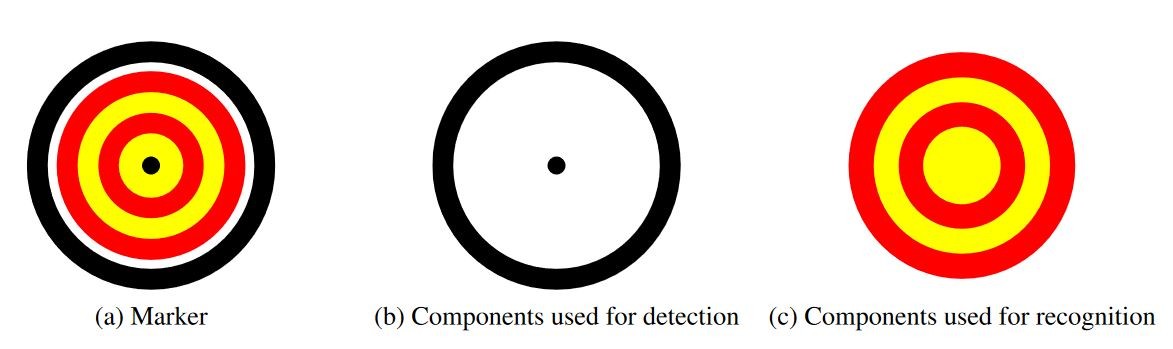}
	\centering
	\caption[Circular markers proposed by \citeauthor{circular-marker}]{\label{circular-marker} Circular markers proposed by \citeauthor{circular-marker} \cite{circular-marker} }
\end{figure}

Circular markers have also been explored and \cite{circular-marker} in fact describes a long-range solution using target-like markers, whose components are shown in figure \ref{circular-marker}.

\subsection{Natural features} \label{natural-feature}
Image feature detectors and descriptors can be used in a wide range of applications, such as image classification or object recognition, but are an essential part of many AR applications. They also are vital for solving the simultaneous localization and mapping problem that will be discussed in section \ref{slam}.

Depending on the specific task, different feature detectors and descriptors can be appropriate, this section will discuss some of the most popular methods.

\subsubsection{Feature detectors}
Feature detectors aim to find interesting key points in an image. Those local features should be invariant to position, rotation and scale changes. They should also be robust to occlusion, noise, illumination change and sufficiently distinct from each other.

Feature detectors are generally classified into 3 categories: single-scale detectors, multi-scale detectors and affine-invariant detectors \cite{features-detection}.

\paragraph{Single-scale detectors}
~\\
The first group, single-scale detectors, can deal with positional and rotational changes of the image. They can also handle noise and illumination changes but they are not designed to cope with scaling issues. They can therefore be helpful for "standard" AR applications that simply need a marker but cannot be used when the same scene has to be recognized from different viewpoints that cause scaling changes. 

\begin{figure}[h]
	\includegraphics[width=11cm]{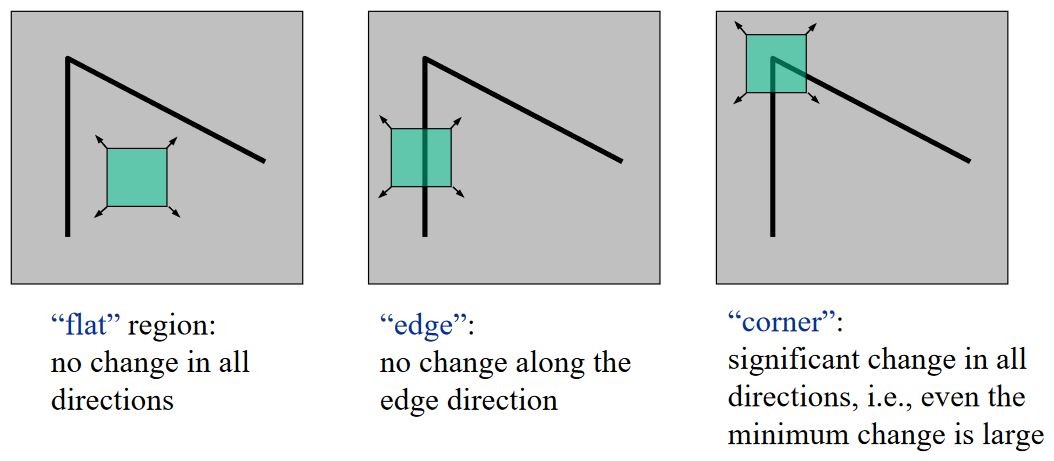}
	\centering
	\caption[Interesting points in an image: effect of translating a windows]{\label{features-good-bad} Interesting points in an image: effect of translating a windows \cite{features-good-bad}}
\end{figure}

A typical example is Harris detector \cite{harris} that detects corners and edges by looking at image gradients (that measure changes in the image's intensity or color). The general intuition behind it is pictured in figure \ref{features-good-bad}.

\begin{figure}[h]
	\includegraphics[width=14cm]{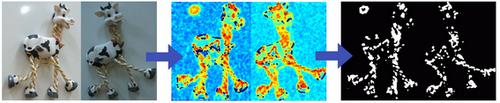}
	\centering
	\caption[The workflow of Harris feature detector]{\label{harris-process} The workflow of Harris feature detector \cite{features-good-bad}}
\end{figure}

\begin{figure}[h]
	\includegraphics[width=7cm]{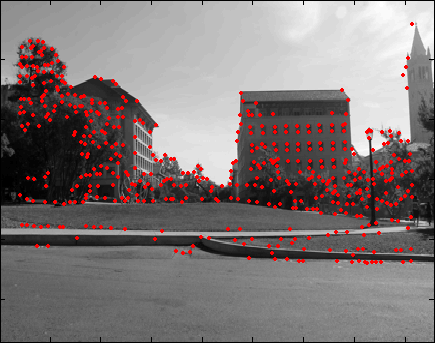}
	\centering
	\caption[Output of Harris detector on an image, with red dots representing detected features]{\label{harris-example} Output of Harris detector on an image, with red dots representing detected features\protect\footnotemark }
\end{figure}

From the source image, a value that depends on the gradient is assigned to each pixel (results are displayed in the second picture of figure \ref{harris-process}, where high values are shown in red). Using a threshold, only interesting points are kept (the white dots from the third picture of the same figure). Finally, only local maxima are kept (the highest values from the "local neighborhood"), those are the key points returned as output, as shown on another example in figure \ref{harris-example}.

\footnotetext{\url{https://inst.eecs.berkeley.edu/~cs194-26/fa14/upload/files/proj7B/cs194-fj/kristen_curry_proj7.2/HearstMining/red5.jpg}} 

\begin{figure}[h]
	\includegraphics[width=10cm]{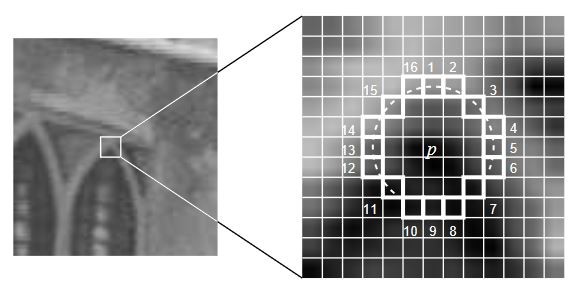}
	\centering
	\caption[A candidate pixel $p$ and the corresponding circle around it using FAST]{\label{fast} A candidate pixel $p$ and the corresponding circle around it using FAST \cite{fast-2}}
\end{figure}

Other methods and adapted versions of the same general idea have been proposed, such as SUSAN \cite{susan} that detects corners with lower-level processing (using a circular mask centered on each pixel of the image to then compare that center pixel's intensity to the rest of the circular area).
FAST (Features from Accelerated Segment Test) \cite{fast-1, fast-2} is a well-known corner detector, that also looks for features by using a circle around a candidate point. This time though, the point $p$ is considered a valid feature if a set of contiguous pixels on the circle are brighter than the candidate (see figure \ref{fast}). To enhance performance, some invalid points can be quickly filtered by only checking a few of those points, e.g. 1, 5, 9 and 13 on the same figure.

\paragraph{Multi-scale detectors}
~\\
Harris detector can easily be adapted to deal with scale changes if the exact deformation is known. Unfortunately, in real world scenarios, scale change is unknown and we need to find ways to detect interesting points at varying scales.

By identifying regions of the image that have properties (e.g. brightness or color) that are different from their surroundings, we form blobs. To find them, a first way is to use Laplacian of Gaussian (LoG) filters that start by smoothing the image with a Gaussian filter (essentially blurring it) then use a Laplacian filter (very noise sensitive, hence the prior smoothing operation) to get those blobs. LoG can therefore be applied for finding a location-specific scale for a region of an image, which means it is possible to automatically select the right scale for that region. \citeauthor{log} proposed a multi-scale approach that does exactly that \cite{log}.

As LoG is computationally expensive, \citeauthor{dog} proposed a more efficient solution that is based on a difference of gaussians (DoG) at different scales \cite{dog}. The input image is successively smoother by a Gaussian filter and resampled. LoG is then essentially approximated by subtracting two successive smoothed images.

\paragraph{Affine-invariant detectors}
~\\

\begin{figure}[h]
	\includegraphics[width=7cm]{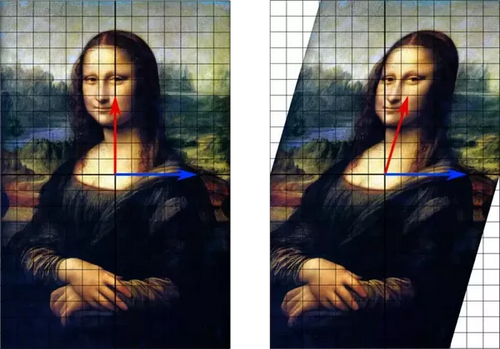}
	\centering
	\caption[Shear mapping and its preservation of parallelism (and therefore perpendicularity)]{\label{shear} Shear mapping and its preservation of parallelism (and therefore perpendicularity)\protect\footnotemark }
\end{figure}

Single-scale detectors discussed previously exhibit invariance to translations and rotations. For their part, multi-scale detectors can also handle uniform scaling and, to some extent, affine invariance (ability to handle shear mapping and non-uniform scaling in addition to previous operations; with shear essentially referring to the possibility of viewing the scene from a different perspective while preserving parallelism, as seen on figure \ref{shear}). 

\footnotetext{\url{https://qph.ec.quoracdn.net/main-qimg-c00a7e8ac5efc42e412270723ec3d459}} 

Affine-invariant detectors go one step further: they are able to handle significant affine transformations. Several existing feature detectors have been extended to handle those perspective issues and methods have been developed, such as the one by \citeauthor{affine-invariant} \cite{affine-invariant}, proposed in \citeyear{affine-invariant}.
The idea of that method is to extract the characteristic shape (as opposed to the characteristic scale discussed before) of the detected feature. The circles are in fact replaced by ellipses with axis lengths that depend on the same gradient-dependent value used in Harris detector \cite{harris} we discussed before.

\begin{figure}[h]
	\includegraphics[width=10cm]{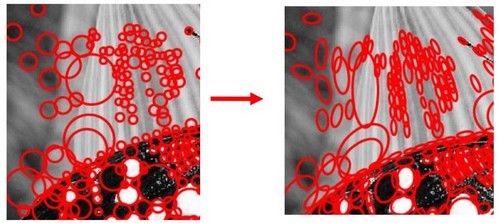}
	\centering
	\caption[\citeauthor{affine-invariant}'s proposal key concept: characteristic shape, with ellipses of different sizes]{\label{affine-invariant} \citeauthor{affine-invariant}'s proposal key concept: characteristic shape, with ellipses of different sizes \cite{affine-invariant-lecture}}
\end{figure}

\subsubsection{Feature descriptors}
Now that feature detectors have been introduced, meaning key points can be detected, it is time to "describe" them so that they can be recognized (and matched). For that purpose, feature descriptors (sometimes referred to as feature extractors), are used. They work by analyzing the feature's surroundings (the pixels around it) and often use techniques described in the previous section.

In \citeyear{sift}, \citeauthor{sift} presented SIFT (Scale-Invariant Feature Transform) \cite{sift}, a detector and descriptor that uses 4 major steps: 
\begin{itemize}
	\item Scale-space extrema detection: use DoG to detect interesting points
	\item Keypoint localization: Determine location and scale, then select keypoints based on their stability
	\item Orientation assignment: Using gradient directions, assign orientations to the keypoints (this step ensures the method is invariant to orientation, scale and location)
	\item Keypoint descriptor: Measure gradients in the region around each keypoint, transform them so that they are more robust to distortion and illumination changes, then store them
\end{itemize}

In \citeyear{surf}, \citeauthor{surf} presented SURF (Speeded Up Robust Features) \cite{surf}, designed as a more efficient version of SIFT.

SIFT and SURF are both patented and use gradients (relatively computationally-expensive and heavy on storage) as descriptors. For these reasons, binary image descriptors have been developed with low power mobile devices in mind. They essentially replace gradient-based encoding by a compact binary string. A few examples of algorithms using them are given below.

BRIEF (Binary Robust Independent Elementary Features) \cite{brief} was introduced in \citeyear{brief} by \citeauthor{brief} and is the first technique that uses binary descriptors. The idea is that, within the local region of a feature, pixels are compared by pairs (chosen using different methods). Depending on how their intensity relates, a binary value is assigned and by concatenating those bits, we get the binary string used as descriptor.

Other well-known examples of binary image descriptors (therefore based on \citeauthor{brief}'s work) include BRISK (Binary Robust Invariant Scalable Keypoints) \cite{brisk}, ORB (Oriented FAST and Rotated BRIEF) \cite{orb} and FREAK (Fast Retina Keypoints) \cite{freak}.

A lot of comparisons between the aforementioned algorithms and their combination (i.e. one of them used for feature detection, another for feature description) have been published. Some are meant to compare these algorithms based on their application to 3D object matching \cite{features-evaluation-3d-object}, pedestrian detection \cite{features-evaluation-pedestrian} or visual tracking \cite{features-evaluation-visual-tracking} (particularly useful for AR and SLAM, that will be discussed in section \ref{slam}). Other focus their analysis on binary descriptors \cite{features-evaluation-binary} or specific evaluation criteria, such as repeatability rate (how stable the features are under different transformations) and information content (how features differ) \cite{features-evaluation-repeatability-information-content}.

\subsubsection{Feature matching}
Once features have been found and described, the next task is to match them in different images. Various possibilities are being used for that purpose, relying on nearest-neighbor approaches or randomized kd-tree forests like the "fast library for approximate nearest neighbors" (FLANN) \cite{flann}. However, those methods are not suitable for binary descriptors, which are typically compared using the Hamming distance (essentially counting the number of mismatching bits between the binary strings $a$ and $b$, that show as 1's after computing $a \oplus b$ (XOR operation)).

\section{SLAM} \label{slam}

The simultaneous localization and mapping (SLAM) problem describes the mapping of an unknown environment by a mobile robot. Without any previous knowledge of its surroundings, the robot therefore has two problems to solve at once: localize itself and map the environment.

SLAM is heavily tied with AR, especially spatial-aware AR, as devices need to get a sense of their environment to successfully integrate the augmented content in their surroundings. This section therefore introduces the basics of the problem.

Even though several researchers were already working on mapping and localization at the time, the structure of the SLAM problem and the coining of the acronym was first presented in \cite{slam-first} by \citeauthor{slam-first} in \citeyear{slam-first}.

\begin{figure}[h]
	\includegraphics[width=13cm]{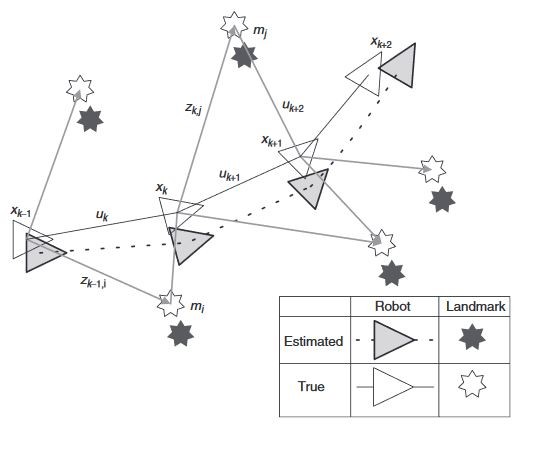}
	\centering
	\caption[The SLAM problem]{\label{slam-def} The SLAM problem, as described in \cite{slam-def} }
\end{figure}

In \cite{slam-def}, \citeauthor{slam-def} give a formal definition of that problem as pictured in figure \ref{slam-def} and described below.
A mobile robot freely moves in an environment while using its sensor to observe unknown landmarks that are assumed to be stationary. 
\\\\
At time $k$, we have:
\begin{equation*}
\begin{cases}
x_k & \text{the vector describing the position/orientation of the robot}\\
u_k & \text{the vector, applied at time } k - 1 \text{ to move the vehicle to } x_k \text{ at time } k\\
m_i & \text{the vector describing the position of the } i^\text{th} \text{ stationary landmark}\\
z_{ik} & \text{the observation, taken from the robot, of the } i^\text{th} \text{ landmark at time } k\\
\end{cases}
\end{equation*}

Motion tracking and natural feature extraction techniques described in previous sections are the usual base tools for solving the SLAM problem, as the landmarks mentioned in the formal definition typically are natural features extracted with algorithms such as those discussed in section \ref{natural-feature}. The position of the camera and the observed landmarks are then inferred but, as noise and inaccuracies are unavoidable, methods have to be developed to cope with them. A very widely-used option is the extended Kalman \cite{kalman} filter (EKF), sometimes with specially trained neural networks \cite{kalman-neural1, kalman-neural2}.

Researchers have been very active trying to solve the problem in the last decades and many approaches have in fact been explored. Some involve multiple robots building the map together while others focus on solving the same problem for hand-held or hand-worn devices (such as PTAM \cite{ptam}).

Recent years have seen compelling commercial products start to appear, with Google Tango\footnote{\url{https://get.google.com/tango/}} or Microsoft Hololens (further discussed in \ref{hololens}) both providing spatial mapping capabilities.

\chapter{Virtual Reality and its use cases} \label{vr}

VR can be used in a very wide range of domains. It is in fact impossible to name all of those fields of applications but a number of examples of use cases and ongoing researches will be discussed in this chapter. Even though entertainment currently is the main driver for VR, we will focus on more "serious" applications.

\section{Healthcare} \label{vr-healthcare}

VR users can experience a sense of presence \cite{vr-app-presence} and that is key to various treatments for mental conditions (e.g. anxiety and specific phobias). In fact, a term has even been coined: virtual reality exposure therapy (VRET). That kind of treatment has been applied to flying phobia \cite{vr-app-flying-phobia}, fear of heights \cite{vr-app-heights-phobia, vr-app-heights-phobia2} or spiders \cite{vr-app-spider-phobia, vr-app-spider-phobia2}, but also to post-traumatic stress disorder (PTSD), with applications to World Trade Center victims \cite{vr-app-ptsd-world-trade-center} or individuals suffering from combat-related PTSD \cite{vr-app-ptsd-war} like US Vietnam \cite{vr-app-ptsd-vietnam-veterans} or Iraq \cite{vr-app-ptsd-iraq} veterans.

But VR applications to medicine are not limited to psychiatric and behavioral 
healthcare, it has also been used in neuropsychology, for the assessment and rehabilitation of disabilities that result from brain injury, memory impairments or attention deficits \cite{vr-app-brain}, while others have focused on stroke rehabilitation \cite{vr-app-stroke, vr-app-stroke2, vr-app-stroke3}. VR can also be used for pain distraction (e.g. during painful interventions) and even chronic pain \cite{vr-app-chronic-pain, vr-app-chronic-pain2}.

\section{Computer-aided design} \label{vr-CAD}

Computer-aided design (CAD) describes the use of computers to create, analyze and optimize designs. It can be applied to many domains, including mechanical, electronic or even orthopedic design activities as well as the architecture, engineering and construction (AEC) industry. In the context of architectural design, the term "Computer-aided architectural design" (CAAD) is often used to describe softwares that accommodate the specific needs of the field.
Those systems can often benefit from computer-generated environments, the term "Virtual engineering" has even been coined and is now widely used in the automotive industry \cite{vr-app-automotive}. Even though they are not always combined with VR, those systems can really benefit from that technology as a strong sense of presence has also been assessed in that context \cite{vr-app-presence-design}.

\begin{figure}[h]
	\includegraphics[width=8cm]{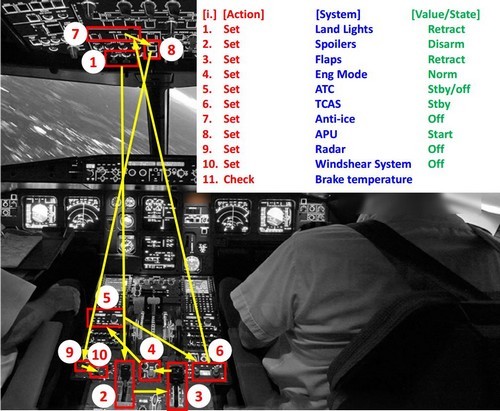}
	\centering
	\caption[Usual actions performed by a pilot in the "after landing" procedure]{\label{cockpit-actions} Usual actions performed by a pilot in the "after landing" procedure \cite{vr-app-cockpit}}
\end{figure}

For example, in the aerospace industry or more specifically in cockpit design, VR can greatly help in evaluating the ergonomics of complex interfaces. A pilot can visualize how the product would look and perform usual actions on it, as pictured in figure \ref{cockpit-actions}. The way he interacts with it can be analyzed and engineers can then validate their virtual prototype.

The technology has also been used for assembly \cite{vr-app-assembly2, vr-app-assembly} and manufacturing \cite{vr-app-manufacturing} simulations, as well as scientific visualization (as part of a product's design phase). The latter category includes fluid simulations such as the airflow visualization tool \cite{vr-app-airflow2, vr-app-airflow} developed by NASA. 

But VR is not limited to single-user experiences and it is in fact a great tool for participatory design \cite{vr-app-participatory, vr-app-participatory2, vr-app-participatory3} where all stakeholders are involved.

\section{Education and training}
The ability to visualize 3D models and environments as if you were part of it can prove very useful in an educational and training contexts, especially when coupled with haptic feedback. Learning by doing is not really different from what one can do through a first-person virtual experience, which means similar learning benefits should be observed. 

For instance, surgical education typically involves animal cadavers or plastic mock-ups but using VR training in that context \cite{vr-app-surgery} was proven to be successful \cite{vr-app-surgery-evaluation} with significant improvements on the trainees' part. Similarly, interesting results were achieved by using the technology for teaching in various domains such as anatomy \cite{vr-app-anatomy}, foreign languages or supply chain. The "fun" aspect of VR probably also helps in keeping users' attention high, which positively affects their performance.

Being able to simulate extreme conditions that would be costly or even risky for users is also a big advantage of virtual reality training, with examples dedicated to firefighters and fire victims \cite{vr-app-firefighter} or even astronauts \cite{vr-app-astronaut, vr-app-astronaut2}.

\section{Culture and tourism}
While VR experiences cannot really be a substitute for real travel, the ability to be transported into places that no longer exist, even if it is only virtual, is attractive.

In fact, VR was already being used for cultural heritage \cite{vr-app-heritage, vr-app-heritage2} prior to the democratization of HMDs for the general public.

\begin{figure}[h]
	\includegraphics[width=10cm]{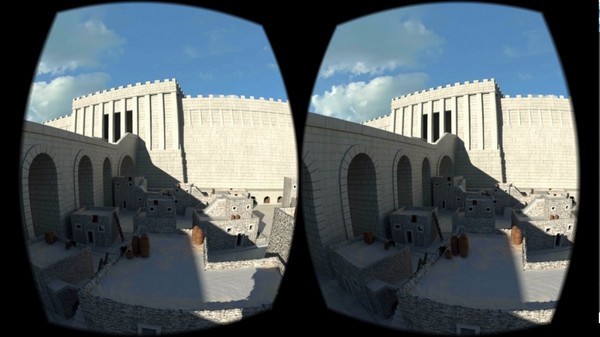}
	\centering
	\caption[Ancient Jerusalem experienced in VR]{\label{ancient-jerusalem} Ancient Jerusalem experienced in VR\protect\footnotemark}
\end{figure}

More recent examples include applications that allows virtual visitors to see ancient Rome\footnote{\url{http://colosseumlives.com/}} or Jerusalem\footnote{\url{https://play.google.com/store/apps/details?id=com.ARE.AncientJerusalemVR}} (a capture from within the application is shown in figure \ref{ancient-jerusalem}).

\footnotetext{\url{https://img.purch.com/h/1400/aHR0cDovL3d3dy5saXZlc2NpZW5jZS5jb20vaW1hZ2VzL2kvMDAwLzA4OS8yMzQvb3JpZ2luYWwvamVydXNhbGVtLXZyLWIuanBn}}

\chapter{Exploring Augmented Reality applications}

As with VR, AR can be used in very different applications. This chapter will discuss a few domains where AR proved helpful. Some examples will voluntarily look similar to what was presented in the previous chapter because, in fact, there is an overlap between AR and VR's fields of application.

\label{ar}
\section{Entertainment}

The entertainment industry provides a major field of application for AR. It is once again impossible not to mention the game Pokémon Go as it truly brought AR to the masses (at a very basic level but it did help popularize the concept). Another common use of AR is in sports broadcasting, as already mentioned in section \ref{ar-def}.

Nevertheless, AR entertainment is not limited to smartphones and TVs. Skemmi\footnote{\url{http://www.skemmi.com/}}, a Belgian company, fully understands that as they are specialized in mass-interactive experiences generally involving AR. A significant part of their work involves cinema events, where everyone in the public can play an AR game broadcasted on the giant screen, using gestural interaction (e.g. by slicing virtual fruits with an arm movement). 

For the launch of Disney's Vaiana, they even enhanced the experience by turning the usual intra-cinema competition to a cinema battle. Two different rooms were in fact competing against each other, trying to paddle as fast as possible so that a pirogue could reach its goal\footnote{\url{https://vimeo.com/194798563}}.

But that kind of experience is not only intended at entertaining people, as psychology studies \cite{skemmi-psycho} have shown that collective experiences have a positive impact on several aspects and can be used to strengthen collective identity or self-esteem for example in the context of teambuilding activities.

\begin{figure}[h]
	\includegraphics[width=9cm]{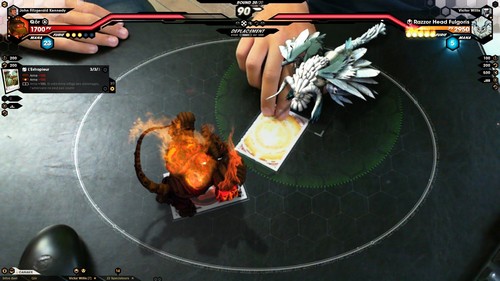}
	\centering
	\caption[Drakerz, an AR collectible card game]{\label{drakerz} Drakerz, an AR collectible card game\protect\footnotemark }
\end{figure}

Other somewhat popular applications of AR in the entertainment industry are card \cite{ar-app-entertainment-card2, ar-app-entertainment-card} and board \cite{ar-app-entertainment-board, ar-app-entertainment-board2, ar-app-entertainment-board-monopoly} games. Commercial examples of the former include Genesis\footnote{\url{http://www.genesisaugmented.com/}} and Drakerz\footnote{\url{https://www.drakerz.com/}}, pictured in figure \ref{drakerz}.

\footnotetext{\url{https://i.ytimg.com/vi/yEaR116swnQ/maxresdefault.jpg}} 

Augmenting those "physical" games with animated characters and other elements can enable very interesting and immersive experiences, by combining the tangible aspect of traditional games with the aesthetics and animations of computer games.

\section{Retail industry, with a practical example} \label{ar-retail}
As part of an entrepreneurial project and together with 4 other students, we developed an AR catalog application. The idea is that potential purchases such as furnitures can sometimes look good but, once bought, they do not always fit in their final environment (wrong dimensions, colors or "style"). 

\begin{figure}[h]
	\includegraphics[width=9cm]{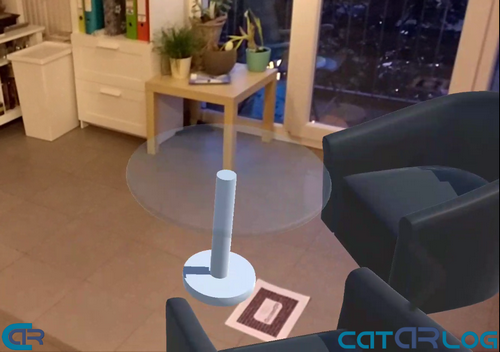}
	\centering
	\caption[A proof-of-concept for an AR catalog application]{\label{catarlog-illustration} A proof-of-concept for an AR catalog application}
\end{figure}

In order to avoid that problem, we propose CatARlog, a mobile application that uses AR to let end customers visualize how objects will look like in their own interior (proof-of-concept shown in figure \ref{catarlog-illustration}). Users simply have to download the store-specific application and place a flyer on the floor that is used as an AR marker. Then, the application lets them choose from a set of models from that store.

Similar applications have been developed, for large companies such as IKEA, Lego or Converse. With CatARlog, we are also targeting SMBs (small and medium size businesses) that generally do not have the same kind of budget.

\begin{figure}[h]
	\includegraphics[width=9cm]{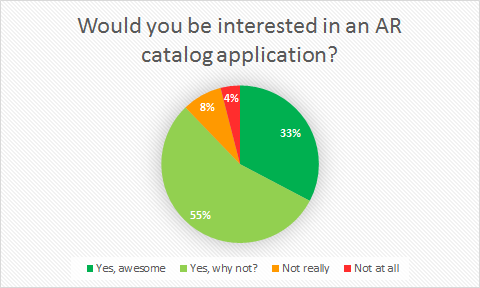}
	\centering
	\caption[Potential users' interest in the AR catalog product]{\label{catarlog-interest} Potential users' interest in the AR catalog product}
\end{figure}

Figure \ref{catarlog-interest} shows that most people are interested in that kind of application (the data comes from a survey we conducted with 50 "random" persons). Additionally, a study from Retail Perceptions\footnote{\url{http://www.retailperceptions.com/2016/10/the-impact-of-augmented-reality-on-retail/}} that questioned more than a thousand US citizens about AR shows, among other things, that 71\% of shoppers would shop at a retailer more often if they offered such a service.
That interest shown by end users is encouraging for the future of AR in retail and a few companies do believe in its potential in that context, such as Augment\footnote{\url{www.augment.com/}} and DigitalBridge\footnote{\url{http://digitalbridge.eu/}}.

\section{Healthcare}

\begin{figure}[h]
	\includegraphics[width=9cm]{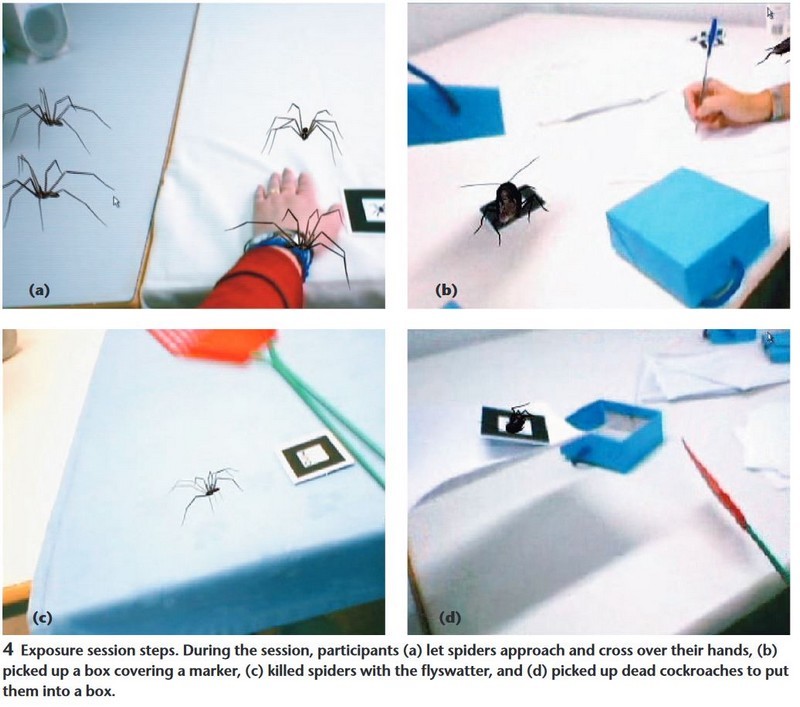}
	\centering
	\caption[Augmented reality exposure therapy for spider and cockroach phobias]{\label{ar-spider-cockroach} Augmented reality exposure therapy for spider and cockroach phobias \cite{ar-app-phobia-spider-cockroach}}
\end{figure}

As with VR, the sense of presence (here enhanced by the fact that the user can actually see his own hands and the real world) can help in exposure therapies for several types of psychological problems, such as spider and cockroach phobias \cite{ar-app-phobia-cockroach, ar-app-phobia-spider-cockroach}. Figure \ref{ar-spider-cockroach} pictures different steps of an AR exposure therapy. More recent work that also includes environmental awareness and does not require markers can be seen at \cite{ar-phobia-youtube}.

Other applications include an AR treatment for phantom limb pain \cite{ar-app-phantom-limb}, overlays for surgeries \cite{ar-app-surgery-laparoscopic, ar-app-surgery-breast-cancer} and post-stroke hand rehabilitation \cite{ar-app-stroke}.
\newpage
\section{Education and training}
\begin{figure}
	\centering
	\begin{subfigure}[b]{6cm}            
		\frame{\includegraphics[width=6cm]{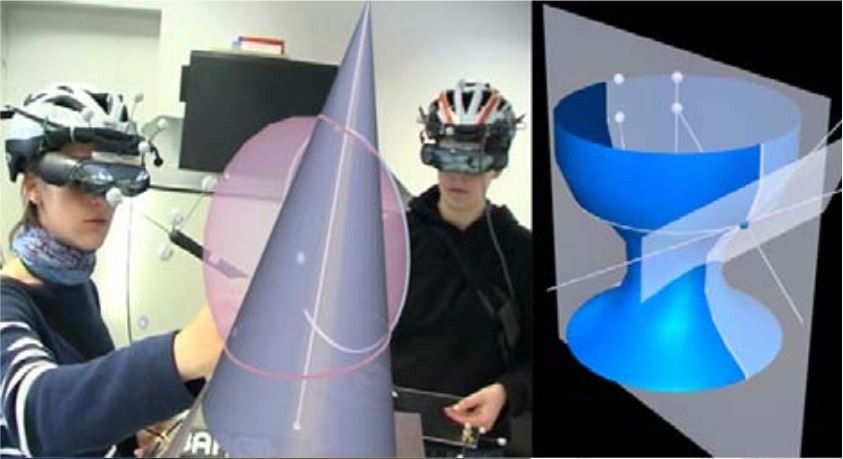}}
		\caption{Collaborative AR experience in the context of maths and geometry teaching \cite{ar-app-teach-maths}}
		\label{ar-app-maths-geometry}
	\end{subfigure}
	\hspace{1cm}
	\begin{subfigure}[b]{7cm}
		\centering
		\frame{\includegraphics[width=7cm]{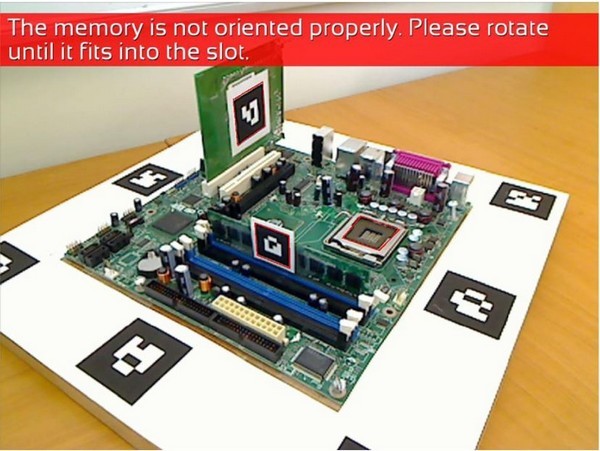}}
		\caption{Motherboard assembly training using AR \cite{ar-app-training-motherboard}}
		\label{ar-app-motherboard}
	\end{subfigure}
	\caption{Examples of AR applications in the context of education and training}
\end{figure}

In education, AR has been used to teach anatomy \cite{ar-app-teach-anatomy}, maths and geometry \cite{ar-app-teach-maths}, engineering \cite{ar-app-teach-engineering} or even astronomy \cite{ar-app-teach-astronomy}. Figure \ref{ar-app-maths-geometry} shows one of these examples in a collaborative teaching context, with a superimposed model of a cone being worked on.

AR has also been employed to train individuals for assembly tasks \cite{ar-app-training-assembly, ar-app-training-assembly2}, including motherboard installation as shown in figure \ref{ar-app-motherboard}, and military operations, such as room-clearing scenarios \cite{ar-app-training-room-cleaning} and in the context of urban terrains \cite{ar-app-training-urban}.

\section{Architecture, engineering and construction}
\begin{figure}[h]
	\includegraphics[width=12cm]{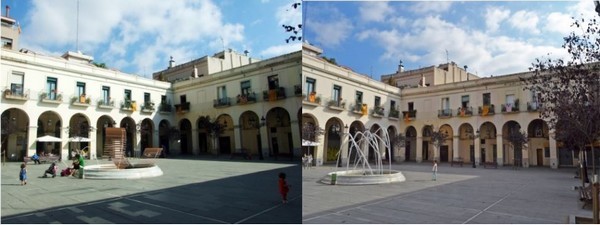}
	\centering
	\caption[Proposed sculptures integrated into their potential context: the middle of Plaza Masadas, Barcelona]{\label{ar-app-aec-square} Proposed sculptures integrated into their potential context: the middle of Plaza Masadas, Barcelona \cite{ar-app-aec-education}}
\end{figure}

As with VR, applications in the AEC industry are numerous, with information overlaid onto buildings using a mobile application \cite{ar-app-aec-mobile} or students' creations superimposed in the middle of a square (see figure \ref{ar-app-aec-square}) to evaluate their design \cite{ar-app-aec-education} .

When maintaining roads or constructing buildings and if the corresponding data/application are available, AR can help visualize underground pipes and subsurface data, as discussed in \cite{ar-app-pipes, ar-app-subsurface, ar-app-subsurface2}. Companies like Bentley\footnote{\url{https://communities.bentley.com/other/old_site_member_blogs/bentley_employees/b/stephanecotes_blog/archive/2012/06/18/augmented-reality-for-subsurface-utilities-further-improving-perception}} have also shown interest in using that kind of subsurface visualization.

\section{Culture and tourism}
Binoculars are widely used in specific touristic locations, allowing visitors to see and zoom in the surroundings for a few minutes by putting a coin in. Those experiences can be augmented \cite{ar-app-tourism-binoculars} by integrating elements on top of the view: some information or pointers and even buildings or structures. As most people (or at least families) now own a smartphone, similar experiences can be offered via mobile devices \cite{ar-app-tourism-mobile2, ar-app-tourism-mobile}.

\begin{figure}[h]
	\includegraphics[width=10cm]{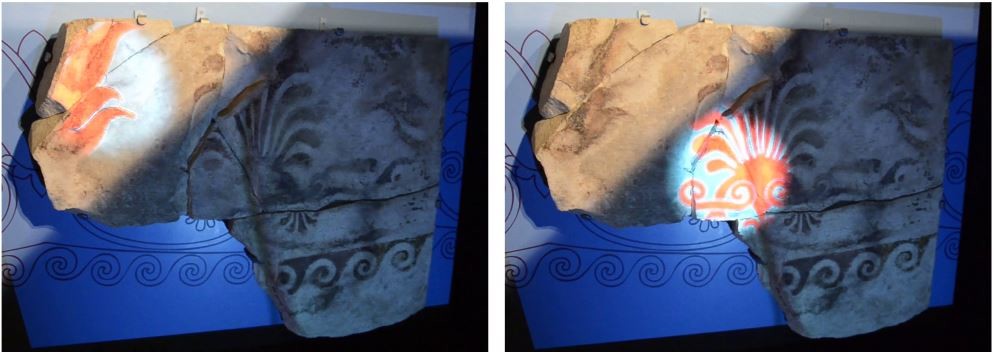}
	\centering
	\caption[Using SAR to colorize archaeological artifacts in a temporary exposition in Rome]{\label{sar-artifact} Using SAR to colorize archaeological artifacts in a temporary exposition in Rome \cite{sar-thesis}}
\end{figure}

Similarly, cultural heritage sites can benefit from the technology, by superimposing
monuments that have since disappeared or virtual inhabitants of the corresponding period \cite{ar-app-heritage2, ar-app-heritage}. Another use of AR in that context has been described in \cite{ar-app-heritage-drawings}, where hard-to-observe animal engravings are highlighted in real time on a smartphone. Similarly, SAR can be used to colorize archaeological artifacts \cite{sar-thesis}, as seen in figure \ref{sar-artifact}. Those artifacts can also be felt if an haptic interface is being coupled with AR, as in \cite{ar-app-artifact-felt}, where the technology was used in museums to create the illusion of feeling objects that are otherwise impossible to touch.

\section{Industrial maintenance and complex tasks} 
For the highly demanding technical work that is often required in industrial maintenance, AR can come to help, for instance by providing additional information superimposed on an object or by displaying a 3D model of the piece being maintained \cite{ar-app-maintenance}. It has been used for helping welders in the automotive industry \cite{ar-app-maintenance-car} as well as the personnel responsible for maintaining a pump \cite{ar-app-maintenance-pump}. 

As operators sometimes need the help of a remote engineer, collaborative AR systems can be a better way of indicating specific pieces than describing them orally. Such remote assistance AR systems were developed in \cite{ar-app-maintenance-collaborative} and \cite{ar-app-maintenance-collaborative2}.

\begin{figure}[h]
	\includegraphics[width=10cm]{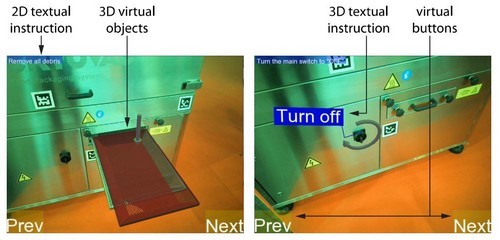}
	\centering
	\caption[Usual virtual objects that are used in an AR maintenance application for the food industry]{\label{ar-app-maintenance} Usual virtual objects that are used in an AR maintenance application for the food industry \cite{ar-app-maintenance-food}}
\end{figure}

Complex assembly tasks typically require the use of manuals containing instructions. Those manuals can be (partly) replaced by AR systems displaying the same instructions and adding world-anchored indications. Those kinds of applications have been developed for domains such as the aerospace \cite{ar-app-maintenance-aerospace} or food \cite{ar-app-maintenance-food} industries. Figure \ref{ar-app-maintenance} shows usual augmented elements used in the latter context.

\chapter{Spatial-aware augmented reality}
\label{mr}
This chapter will present what we chose to call spatial-aware augmented reality (spatial-mapped AR or surroundings-aware AR would have been valid names too).
This subset of AR has a better understanding of the environment than traditional AR, it does not simply add augmented content "blindly", neither does it rely on (fiducial) markers to place that content.
\section{Microsoft Hololens}
\begin{figure}[h]
	\includegraphics[width=6cm]{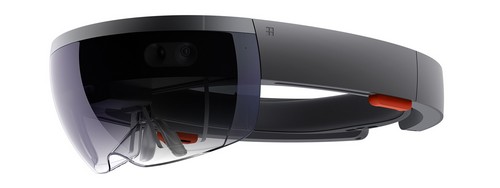}
	\centering
	\caption[Microsoft Hololens, a spatial-aware AR HMD]{\label{hololens} Microsoft Hololens, a spatial-aware AR HMD\protect\footnotemark }
\end{figure}

An example of a device enabling spatial-aware experiences is Microsoft Hololens. The real innovation with this headset is that it combines several advanced technologies into a single, autonomous and portable device (shown in figure \ref{hololens}).
\footnotetext{\url{https://compass-ssl.surface.com/assets/f5/2a/f52a1f76-0640-4a37-a650-51b0902f8427.jpg}} 
\subsection{The hardware} \label{hololens-hardware-section}
\begin{figure}[h]
	\includegraphics[width=8cm]{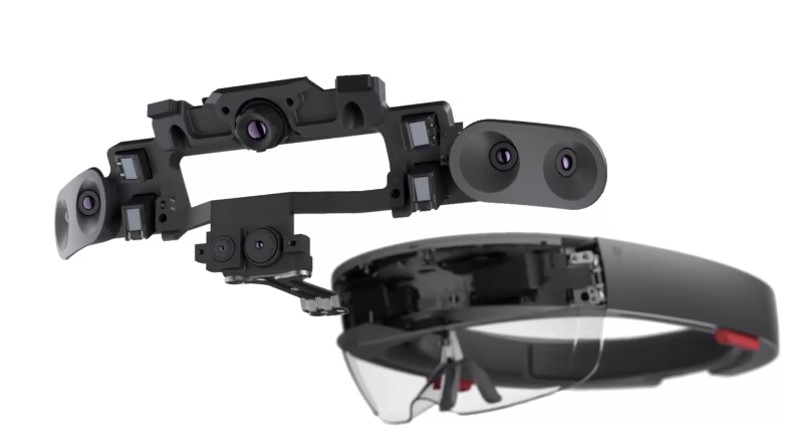}
	\centering
	\caption[Microsoft Hololens cameras and sensors]{\label{hololens-cameras} Microsoft Hololens cameras and sensors\protect\footnotemark }
\end{figure}

The Hololens uses see-through lenses and 2 light engines to project the augmented content. It automatically calibrates pupillary distance, has a "holographic resolution" of 2.3M total light points and a "holographic density" of more than 2.5k light points per radian. In order to scan the environment, it uses 4 dedicated cameras, in addition to the depth camera and the 2MP photo/video camera (see figure \ref{hololens-cameras}). It also has an IMU (inertial measurement unit, to track head movements), 4 microphones, an ambient light sensor and a spatial sound system. More information can be found at \cite{hololens-hardware, hololens-hardware-2}. As discussed in section \ref{inertial}, an IMU alone is not sufficient to track something continuously. Unfortunately, Microsoft did not reveal the method they used to remove unavoidable drifts but they did explain \cite{hololens-science-within} that spatial mapping was using natural feature extraction (see section \ref{natural-feature}) so it is very likely that positional tracking is corrected that way.

\footnotetext{\url{https://az835927.vo.msecnd.net/sites/mixed-reality/Resources/images/Sensor_bar.jpg}} 

To process sensors data and camera feeds, Microsoft created a custom chip called the Holographic Processing Unit (HPU). More information on that particular component of the device can be found in \cite{hololens-hpu}.

\subsection{"Mixed reality" experiences}
Microsoft and lots of websites describe the Hololens as a mixed reality device. While this is not fundamentally false, it leads to confusion as MR is thus often misunderstood as "AR with real world understanding and anchoring". As explained in chapter \ref{def-taxonomy}, MR is a superset of AR, therefore neither a subset nor something different from AR. In fact, the Hololens is an AR device and Hololens applications are AR experiences.

That being said, spatial-aware AR has many current and potential applications. Just like VR, entertainment and more specifically video games have embraced the technology. But there are also lots of industrial applications, most of whom are yet to be explored. 

\begin{figure}[h]
	\includegraphics[width=10cm]{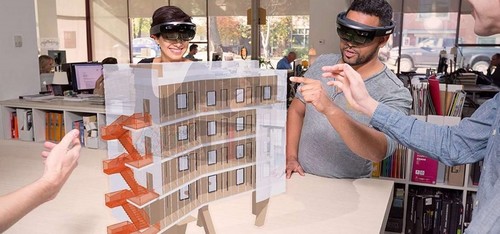}
	\centering
	\caption[Sketchup, real-time collaboration for the AEC industry]{\label{hololens-sketchup} Sketchup, real-time collaboration for the AEC industry\protect\footnotemark }
\end{figure}

For example, the AEC (architecture, engineering and construction) industry could benefit from such a technology. They already transitioned from 2D hand drawings to 2D digital plans, then to 3D models. But current 3D models are currently stuck behind a 2D screen. What if they could be integrated into the real world (say on a table) and designers could collaborate on it in real time (potentially even remotely)? Figure \ref{hololens-sketchup} shows Sketchup for Hololens, an example of a commercial AEC application from Trimble \cite{hololens-trimble-paper}.

\footnotetext{\url{https://img.reality.news/img/74/25/63614387056309/0/trimble-releases-sketchup-viewer-first-commercial-hololens-application-windows-store.1280x600.jpg}} 

The technology can also be used in maintenance (or more generally when expertise is needed) where lengthy manuals can be replaced by an application that directly integrates instructions and indicators onto the corresponding parts of the object that is being worked on. Combined with things like voice recognition, it is also possible to achieve a quick and hands-free experience that improves productivity. 

An example of such an application (that also includes remote collaboration) for "pipe maintenance" can be seen in \cite{hololens-pipe}. Lots of other scenarios in the retail industry or in education benefiting from the Hololens and similar devices can be imagined but will not be discussed here.

\section{Other devices}

In this section, we will talk about competitors to the Hololens, with a few devices able to deliver similar experiences (or believed to be).

\subsection{Direct competitors} \label{competitors}
\subsubsection{Meta}
Meta\footnote{\url{https://www.metavision.com/}} is an American company that developed Meta 1 and, since December 2016, Meta 2. They both are see-through AR HMDs equipped with a 3D camera, capable of recognizing gestures and superimposing holograms into the real world.
While Meta 1 was somewhat limited (e.g. with a 23-degree field of view), Meta 2 is aiming at competing with the Hololens. Its main characteristics are:
\begin{itemize}
	\item 90-degree field of view (much better than the Hololens' speculated 30 degrees\footnote{\url{http://doc-ok.org/?p=1223}})
	\item 2560x1440 display resolution
	\item 720p front-facing camera
	\item hand tracking capabilities
	\item tethered (requires a Windows 8+ PC)
	\item can be preordered for \$949
\end{itemize}
\subsubsection{Magic Leap}
Magic Leap is a very secretive American startup that was able to raise \$1.4 billion from investors including major companies such as Google, Qualcomm or Alibaba Group.
Since then, the company has been working in "stealth mode" and whether their product achieves what can be seen in their concept video\footnote{\url{https://www.youtube.com/watch?v=kPMHcanq0xM}} is currently unknown (so are the hardware specifications and the price). We only know they want to create a standalone device with a very wide field of view.

\subsubsection{DAQRI}
DAQRI is another American company selling AR headsets. It should also be mentioned that in 2015 they acquired ARToolWorks, the company that initially released ARToolKit. They first started with an AR helmet (to meet the safety requirements in specific industries) with special components such as thermal sensors but are now also selling "standard" AR glasses for \$4995. Both of those devices are using a RealSense 3D camera (from Intel, based on technologies from the Belgian company SoftKinetic) and are offering a 44-degree field of view.

\subsubsection{Others}
Other devices worth mentioning are castAR and ODG's R-8 and R-9. Those see-through glasses are yet to be released but promise very high field of view.

\subsection{Other versions of the Hololens}
A "Hololens v2" was meant to be released (an improved and consumer-ready version) but the project has been sidelined. Instead, the focus will be on a "Hololens v3" planned for 2019, that will most likely introduce major improvements.
In addition to that, Microsoft revealed they partnered with several companies (Acer, Asus, Dell, HP and Lenovo) to create tethered VR headsets with cameras. As of now, it hasn't been clarified whether those cameras will enable video see-through AR or "simply" provide real world information. Those headsets should be available for sale later this year (2017) for \$300.

\section{A practical example: HoloEscape}
As an illustration of what spatial-aware AR can achieve and in order to validate the practical usability of such applications, a game has been developed as part of the present master's thesis.
HoloEscape runs on the Hololens, involves spatial mapping and basic understanding as well as gestural interaction and gaze input.

It has been presented during the "Printemps des Sciences" on March 25-26, 2017 (a public event in Mons, Belgium). As end-user (informal) feedback was intended to be gathered during that event and given that the public mainly consisted of families with children, developing a game seemed like an obvious choice and was in fact confirmed by the constant line of people wanting to try it.

The device (Microsoft Hololens) was chosen because it is so innovative: being able to see the real world through autonomous glasses capable of superimposing holograms anchored in the environment (occluding the models if necessary thanks to the spatial mapping capabilities) had never been achieved before. In addition to that, as the device was fairly new (it was not even purchasable in Belgium when this master's thesis started), the ability to develop software for it was a unique chance that could not be missed.

\subsection{Game description} \label{game-desc}
The game itself is a "reversed tower defense" where the player controls a virtual ball using gaze input. His goal is to make it reach the end of a holographic road (displayed on the floor) without touching electric walls. Hostile turrets can be added by positioning printed images (which means spectators can effectively take part in the game). Once in position, those turrets will constantly shoot at the player that needs to dodge the resulting laser beams.

When the ball touches a wall or is hit by a laser beam, the player looses one life (out of 3). To enhance spectators' experience, a few mobile mini-games were developed and allowed players to affect the Hololens gameplay. For instance, getting bad scores on a Flappy Bird clone led to a bigger ball (harder to control) and faster beams whereas winning on a Breakout clone could lead to an "unbreakable" ball and slower beams.

\begin{figure}[h]
	\includegraphics[width=10cm]{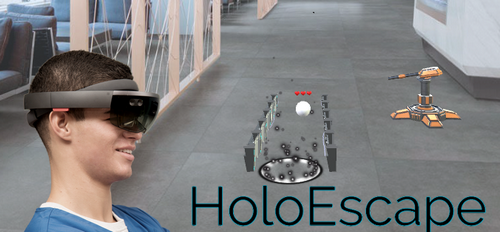}
	\centering
	\caption{\label{holoescape-illustration} HoloEscape: Game concept}
\end{figure}

Figure \ref{holoescape-illustration} helps clarifying the concept (a video is also available at \cite{holoescape-video}).
\subsection{Technologies}
The main technologies the project is built on will be introduced in this section. Potential alternatives will be discussed and choices will be explained. That includes the game engine as well as features such as spatial mapping and gaze input. Finally, the tool used to recognize printed images will be presented and evaluated.
\subsubsection{Unity}
Unity\footnote{\url{http://unity3d.com/}} is a leading game engine, used by millions of people every day (according to their website). It can produce games for many desktop, mobile and console platforms and is recommended by Microsoft for building 3D Hololens applications. Other alternatives include UrhoSharp\footnote{\url{https://developer.xamarin.com/guides/cross-platform/urho/introduction/}} (.NET bindings of Urho3D\footnote{\url{https://urho3d.github.io/}}) and even direct DirectX projects.

Because it is recommended and widely used, Unity has been chosen. It should however be mentioned that the process for building Hololens applications is a bit unusual. Whereas "standard" platforms can be targeted via a simple compilation in Unity's editor, building for Hololens requires an additional step as Unity is only able to create an intermediary Visual Studio\footnote{\url{https://www.visualstudio.com/}} solution. That solution can then be compiled to the actual application (inside Visual Studio). The need to rely on two different compilers leads to issues that will be further discussed in section \ref{issues}.

\subsubsection{Spatial Mapping}
\begin{figure}[h]
	\includegraphics[width=10cm]{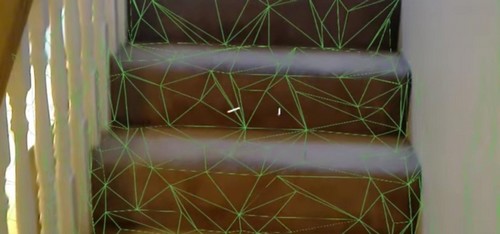}
	\centering
	\caption[An example of a spatial mapping mesh for some stairs]{\label{hololens-spatial-mapping-mesh} An example of a spatial mapping mesh for some stairs\protect\footnotemark }
\end{figure}

Using the hardware described in section \ref{hololens-hardware-section}, the Hololens is able to reconstruct triangle meshes representing its understanding of the environment (example shown in figure \ref{hololens-spatial-mapping-mesh}).
\footnotetext{\url{https://img.reality.news/img/29/16/63622861744313/0/video-space-sound-used-help-visually-impaired-navigate-with-hololens.1280x600.jpg}} 

In the "Holographic Academy"\footnote{\url{https://developer.microsoft.com/en-us/windows/mixed-reality/academy}} (containing a few tutorials explaining how to use the device with Unity) and in the "HoloToolkit for Unity"\footnote{\url{https://github.com/Microsoft/HoloToolkit-Unity}} (a GitHub repository gathering lots of scripts and components related to Hololens development), Microsoft shows how to access and process that spatial mapping data. 

The components they provide come with a few parameters to address the needs of the developers (mesh "resolution", "scanning zone", delay between spatial mapping updates, etc).

In the context of HoloEscape, the only thing we need to extract from that mesh is the floor (to place holographic roads and the ball). To achieve that goal, a possibility is to use HoloToolkit's \textit{SurfaceMeshesToPlanes}\footnote{\url{https://github.com/Microsoft/HoloToolkit-Unity/blob/master/Assets/HoloToolkit/SpatialMapping/Scripts/SpatialProcessing/SurfaceMeshesToPlanes.cs}} that will "convert" the meshes to planes. In our case, we are only interested in horizontal planes as they are good candidates for the floor. It has been chosen to then select the lowest of those candidates that is close enough to the user. In most cases, this will correctly pick the plane that corresponds to the floor of the room the user is in. The only issue I encountered is when the user looks at a contiguous room through a glass during the scanning process. If that other room is lower than the room the user is in, the wrong plane will be picked. As those condition are quite far-fetched and easily avoidable by not looking at that other room, no specific logic was implemented to prevent the issue from happening.

\begin{figure}[h]
	\includegraphics[width=5cm]{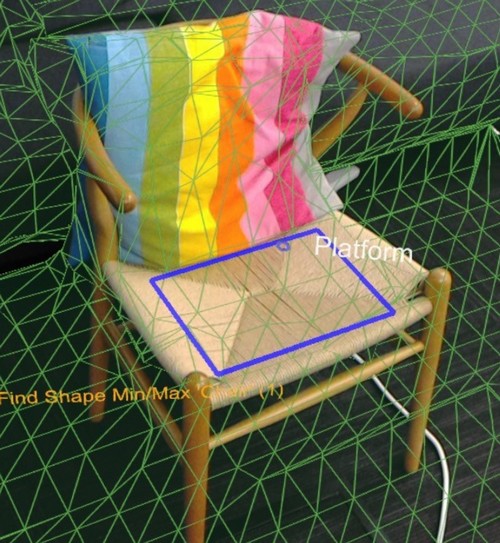}
	\centering
	\caption[Spatial understanding capabilities]{\label{hololens-spatial-understanding} Spatial understanding capabilities\protect\footnotemark }
\end{figure}

Another possibility would have been to use Spatial Understanding scripts\footnote{\url{https://github.com/Microsoft/HoloToolkit-Unity/tree/master/Assets/HoloToolkit/SpatialUnderstanding}}. Those components go beyond a simple "meshes to planes" conversion, they provide a better and higher-level understanding of the environment (example shown on figure \ref{hololens-spatial-understanding}).
\footnotetext{\url{https://az835927.vo.msecnd.net/sites/mixed-reality/Resources/images/SU_ShapeQuery.jpg}} 

These scripts were initially developed by Asobo Studios\footnote{\url{http://www.asobostudio.com/}}, a video game development company based in Bordeaux, France. They had to develop those features for their games\footnote{\url{http://www.asobostudio.com/games\#filter=.hololens}} but then decided to share the corresponding code. Their work is available through a DLL (compiled from their codebase in C++) exposed to Unity in the HoloToolkit.

\begin{figure}[h]
	\includegraphics[width=10cm]{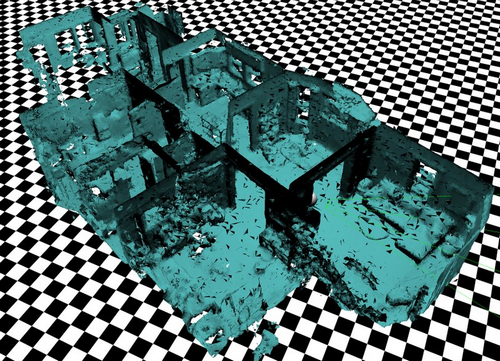}
	\centering
	\caption[A mesh obtained from a mixed reality capture that corresponds to several rooms]{\label{hololens-mixed-capture} A mesh obtained from a mixed reality capture that corresponds to several rooms\protect\footnotemark }
\end{figure}

One should note that it is also possible (using the device portal available through a local website) to obtain a complete mesh for an entire "mixed reality capture", as seen in figure \ref{hololens-mixed-capture}.
\footnotetext{\url{http://www.sharpgis.net/image.axd?picture=image_131.png}} 

\subsubsection{Gaze input}
\begin{figure}[h]
	\includegraphics[width=7cm]{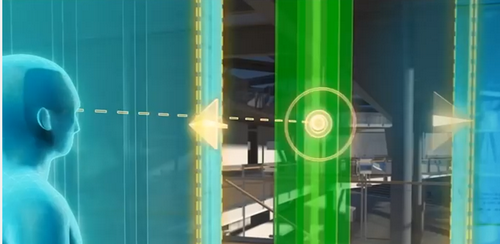}
	\centering
	\caption[Gaze input using of the camera's forward vector]{\label{hololens-gaze} Gaze input using of the camera's forward vector\protect\footnotemark }
\end{figure}

As explained in section \ref{game-desc}, gaze input is used to control the virtual ball in our game. We already know that the IMU (see section \ref{hololens-hardware-section}) is tracking head movements but we have yet to explain how that data can be used by developers. That part is in fact very simple: the main camera of the Unity scene is "mapped" to head movements, its forward vector therefore indicates where the user is looking. It is then fast forward to search for an intersection between a target surface and the semi-straight line produced by that vector (see figure \ref{hololens-gaze}).
\footnotetext{\url{https://abhijitjana.files.wordpress.com/2016/05/image31.png}} 

As with any kind of sensor, there is noise in the raw data received from the IMU which can cause jitter when displaying the cursor. Once again, the HoloToolkit proves to be useful as it contains a \textit{GazeStabilizer} script that can be used to smooth that data (different smoothing filters can of course be applied if needed).

\subsubsection{Image recognition and tracking}
In order to add turrets, one has to place printed images on the floor and "show" them to the camera. A very well-known, open-source and widely used possibility is ARToolkit\footnote{\url{https://artoolkit.org/}}. Unfortunately, it does not currently support Hololens even though \citeauthor{artoolkit-hololens} announced they successfully integrated ARToolkit 5 with it \cite{artoolkit-hololens}.

\begin{figure}[h]
	\includegraphics[width=13cm]{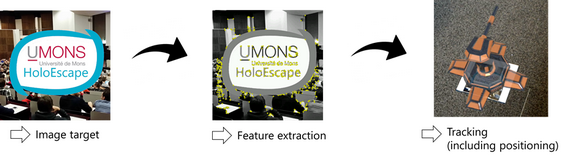}
	\centering
	\caption{\label{vuforia-process} Vuforia's image tracking principle}
\end{figure}

Vuforia\footnote{\url{https://vuforia.com/}} officially supports Hololens and has therefore been chosen. It is capable of recognizing and tracking 3D objects (to some extent) but also image targets (using natural feature extraction, see section \ref{natural-feature}). The process' principle is shown in figure \ref{vuforia-process} (many AR solutions are based on the same principle). The quality of the tracking in fact highly depends on the features (sharp, spiked, chiseled details in the image) that Vuforia was able to extract from the original image. Advices on how to choose and improve those target images can be found in \cite{vuforia-features}. The chosen image that contains the university's logo is not perfect but is still rated with 5 stars out of 5 by Vuforia's target manager, because as seen in the central picture of figure \ref{vuforia-process}, enough features were found in the background image (mostly on persons) and around the letters.

\begin{figure}[h]
	\includegraphics[width=8cm]{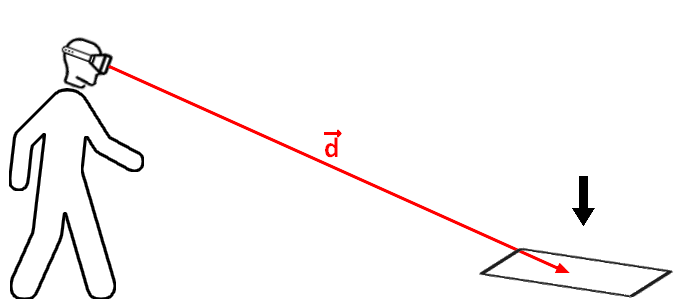}
	\centering
	\caption{\label{hololens-walking} The testing procedure to validate Vuforia's image recognition feature}
\end{figure}

With the aim of validating the image recognition solution, a test was undertaken (by the author) and is explained below.
The user places a printed image on the floor, then moves a few meters back.
With the Hololens (running an application developed for that purpose) on his head, the user slowly walks towards the image. In the process, he makes sure to keep it in sight (a cursor is displayed, much like the yellow circle on figure \ref{hololens-gaze}, and always stays on the printed picture). The situation is pictured on figure \ref{hololens-walking}.

When the user comes close enough to the image, the camera recognizes it. The distance between the device and the paper is measured (the length of $\color{red}\overrightarrow d$ on the figure).

\begin{figure}[h]
	\includegraphics[width=10cm]{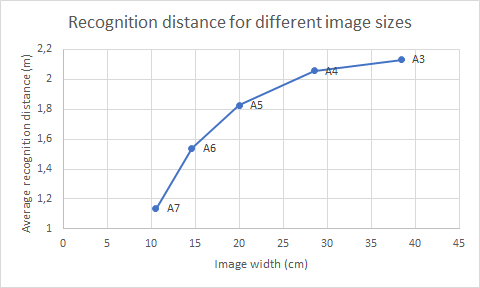}
	\centering
	\caption{\label{vuforia-distance-plot} Effect of a varying image size on Vuforia (for Hololens) "recognition distance" - scatter plot}
\end{figure}

That procedure was tested with a photograph that is classified as an excellent tracking candidate by Vuforia's target manager (5 stars out of 5, indicating that many features have been detected and are spread across the image). The picture was printed in 5 different sizes (labeled with their approximate ISO paper format equivalent) to observe the effect of changing the target's size on the "recognition distance". The corresponding results are displayed in table \ref{vuforia-distance-table} and figure \ref{vuforia-distance-plot}.
Note that A7 is so small it requires the user to bend a little.

\begin{table}[h]
	\begin{center}
\begin{tabular}{|c|c|c|c|}
	\hline
	Label & Image dimensions (cm) & Avg. distance (m) & Std. deviation (m)\\
	\hline
	A3 &	38.4 x 26.9 	&2.1320294	&0.029664227		\\
	A4 &	28.5 x 20		&2.0558848	&0.016190217		\\
	A5 &	20 x 14			&1.8263826	&0.029573648		\\
	A6 &	14.6 x 10.2		&1.5392008	&0.003787432		\\
	A7 &	10.5 x 7.3		&1.135628	&0.017326755		\\
	\hline
\end{tabular}
\end{center}
\caption{\label{vuforia-distance-table} Effect of a varying image size on Vuforia (for Hololens) "recognition distance" - results}
\end{table}

As shown by those results and as expected, increasing the size of the image helps making it easier to recognize, but that recognizability doesn't increase linearly. Another surprising fact is that the standard deviation is relatively low (less than 2\% for all sizes) which means the results were particularly stable.

\subsection{Issues and validation}\label{issues}
Rather than using a virtual ball, the initial goal was to integrate Sphero\footnote{\url{http://www.sphero.com/}}, a robotic ball controlled via Bluetooth, into the game. The technical difficulties that prevented that from happening will be discussed here.

\subsubsection{Connecting Hololens to Sphero directly}

Several factors are causing issues when trying to make Hololens and Sphero communicate: 
\begin{itemize}
	\item Hololens runs on Windows 10, thus able to run "Windows 10 applications" (technically UWP for Universal Windows Platform). Older APIs are therefore not available on the device
	\item The Sphero SDK for Unity only works on Android and iOS which means the Windows SDK has to be used. That SDK hasn't been updated in years (last commit was made in 2013 and it does not work on recent versions of Windows 10 that are required for the Hololens)
	\item Unity uses a modified version of Mono\footnote{\url{http://www.mono-project.com/}} that roughly corresponds to .NET 3.5 (~10 years old). Newer APIs are therefore not available in Unity (and some obsolete APIs used in Unity's Mono cannot be used on the Hololens)
\end{itemize}

Even though the Hololens is capable of Bluetooth connectivity, those issues mean that connecting the device with Sphero is not as simple as it should be.
Several forks of the official Sphero SDK for Windows do exist, some of them are relatively up-to-date and, in a blog post \cite{sphero-blog-mtaulty}, \citeauthor{sphero-blog-mtaulty} even managed to make it work on the Hololens (using SoftPlay's reworked version\footnote{\url{https://github.com/SoftPlay/SpheroWindows/}} of the Sphero SDK).

Unfortunately, that experiment as well as the aforementioned forks are not targeting 3D Unity applications but "only" standard UWP. What \citeauthor{sphero-blog-mtaulty} did is not replicable and building the corresponding library (RobotKit.dll from SoftPlay's GitHub) did not work as exceptions were thrown when trying to retrieve the Bluetooth RFCOMM service that correspond to the Sphero. At the time of writing, no solution to that problem has been found, neither by myself nor by \citeauthor{sphero-blog-mtaulty}.

\subsubsection{Using a relay}
\begin{figure}[h]
	\includegraphics[width=4cm]{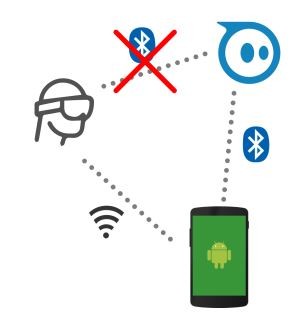}
	\centering
	\caption{\label{sphero-relay} Android application as a relay using WebSockets}
\end{figure}

As a direct connection did not seem possible, another approach was tried: an Android application acting as a relay. In fact, the Sphero SDK for Android works fine so the ball could be controlled by an Android application that communicates with the Hololens using WebSockets (corresponding diagram shown in figure \ref{sphero-relay}). The exchanged messages use FlatBuffers\footnote{\url{https://google.github.io/flatbuffers/}} to format the data in binary buffers (different message types are declared, such as STATUS\_REQUEST or DIRECTIVE, and corresponding data structures are defined) to be as efficient as possible.

However, using WebSockets appeared to be a bit slow for real-time control and tracking of the ball (trusting Sphero's own tracking), as shown by the tests that were carried out. The corresponding procedure is explained below.

As we are interested in bidirectional communication and because it would be very difficult to synchronize the Hololens with the Android application, we measured the round-trip time (RTT) i.e. the total time needed for a message sent from the Hololens to return to it. 

\begin{figure}[h]
	\includegraphics[width=13cm]{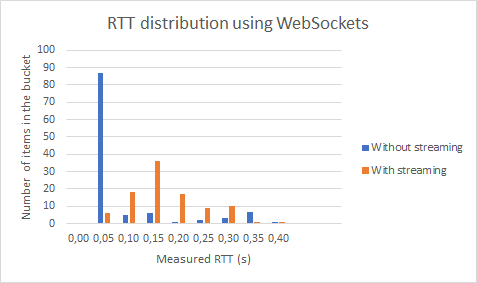}
	\centering
	\caption{\label{websockets-rtt-plot} Round-trip times using WebSockets, with and without streaming - plot}
\end{figure}

To be more precise, the Hololens application starts counting when it sends a message to the phone. As the Android application receives that message, it replies with a predefined response instantly. When the Hololens receives that reply, it stops the counter and stores the result. It then waits for a second and the process can start again (117 complete "trips"). In order to limit interferences, the Wi-Fi used for that communication is created on a laptop, with no other devices connected to it. As only one person at a time can wear the Hololens, the setup has also been tested with live streaming enabled.
The corresponding results are given in table \ref{websockets-rtt-table} and figure \ref{websockets-rtt-plot}. Note that the testing procedure does not take the bluetooth transmission (from the phone to the ball) into account, but it is considered negligible.

\begin{table}[h]
	\begin{center}
		\begin{tabular}{|c|c|c|}
			\hline
			& Without streaming & With streaming\\
			\hline
			Number of "trips"&	117&117\\
			Minimum RTT (s)&	0.0165062&0.03204155	\\
			Maximum RTT (s)&	2.780609&4.49617	\\
			Average RTT (s)&	0.125426211&0.306582943	\\
			Standard deviation (s)&	0.322630263&0.558482321\\
			Median RTT (s)&	0.03330231&0.134491	\\
			Number of RTTs > 0.4s&	5&19\\
			\hline
		\end{tabular}
	\end{center}
	\caption{\label{websockets-rtt-table} Round-trip times using WebSockets, with and without streaming - results}
\end{table}

As there was a significant amount of outliers, two metrics were added: the median RTT and the number of RTT above 0.4s (that are not shown on the plot for readability).
Those results show that without streaming, the solution could be viable but would need some extra precautions to handle "silent periods" as some messages took way too long. A potential solution could be to run a timer on the Android application that tells the ball to stop moving when no directive has been received for a certain delay. On the other hand, we can see the effect of having streaming enabled: the number of RTTs above 0.4s increases (to reach a bit more than 16\% of all samples) and that is hardly acceptable, even though the median value is reasonable.

\subsubsection{Tracking the ball} \label{sphero-tracking}
\begin{figure}[h]
	\includegraphics[width=5cm]{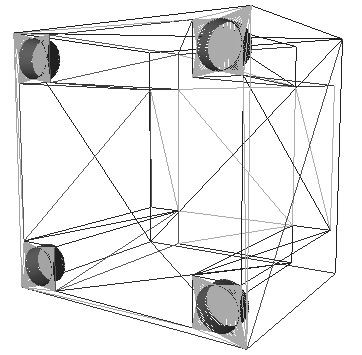}
	\centering
	\caption{\label{sphero-box} Wireframe model of a 3D printed box for the Sphero ball}
\end{figure}

Despite those mixed results, coupling a somewhat slow WebSockets relay with an effective tracking system could still yield interesting results.
Unfortunately, there is currently no way of accessing the raw depth data the Hololens processes, which means that tracking dynamic objects is difficult and less accurate than it could be with that data. Consequently, the only option to track the ball is the front RGB camera (red-green-blue, a standard color camera).
Vuforia is capable of tracking some 3D objects but Sphero is a rolling sphere,  almost entirely white (that can only be lit by a solid color) so it is not suitable in itself as there is no interest point to track. To overcome that issue, a cube with bearing balls has been 3D printed (see figure \ref{sphero-box}).

The idea was to put it around the Sphero ball (made with a translucent material so that the ball's color is reflected) with "trackable" images on some of its sides. That solution would provide a relatively unstable tracking (issues when the user would move his head and stop gazing at the ball/cube) and, as the size of the tracked images would be comparable to the "A7" image from table \ref{vuforia-distance-table}, similar tracking distances would be observed (the user would have to stay close to the ball).
More importantly, it would not allow streaming as only one process can have access to the Hololens' front (and only accessible) camera at any time\footnote{\url{https://developer.microsoft.com/en-us/windows/mixed-reality/mixed_reality_capture_for_developers\#simultaneous_mrc_limitations}}.
As this project was to be presented in public events, such a limitation was inappropriate and the virtual ball was the only viable solution.
\chapter{Current limitations and foreseeable prospects}
\label{limitations-prospects}
VR enables a wide range of applications and can deeply change a set of human endeavors, as seen in chapter \ref{vr}, through short-term experiences. The necessary equipment is relatively mature, whereas AR's potential (in particular through head-worn displays) currently is more restrained by the underlying technological needs. However, once these problems will be solved, AR could be integrated in our daily lives. This chapter will discuss current limitations and future prospects, for both fields.

\section{Displays}

In their famous and previously cited paper \cite{milgram-continuum}, \citeauthor{milgram-continuum} identified the needs for optical see-through displays: \textquote{\itshape accurate and precise, low latency body and head tracking, accurate and precise calibration and viewpoint matching, adequate field of view, [...] a snug (no-slip) but comfortable and preferably untethered head-mount}.

While the Hololens' head tracking and viewpoint matching are already compelling, inside an untethered (autonomous and wireless) device, the field of view still is a major issue. As previously discussed (section \ref{competitors}), competitors announced much wider field of views but whether their promises will be delivered remains to be seen.

As already mentioned in section \ref{displays}, retinal displays could very well be the future as they could potentially combine more portability, a very wide field of view and less eye-tiredness.
The VR industry could benefit from it but it would truly be a game changer for AR as small field of views and unsuitability for outdoor use are the usual issues for see-through devices. Reducing eye-tiredness is also key to long term (maybe even permanent) use.

Many people wear glasses constantly so size and weight should not really be an issue but if tinier displays are desired, contact lenses are also a possibility.
In fact, in \citeyear{lenses-patent}, Samsung patented \cite{lenses-patent} such AR lenses in South Korea. The patent application describes the lenses as equipped with a camera, an antenna and several sensors (according to a brief translation from \cite{lenses-article}).

The patent shows that it might be a work in progress but lenses are so small that compelling experiences look out of reach in the near future. Even though simple overlays could potentially be superimposed at some point, with the computing part handled by a smartphone, latency would probably be a problem. Another issue to solve is power supply, how would lenses get enough energy to run the hardware?

\section{Computing resources and sensors} \label{limitations-computing}

As the hardware gets shorter, more powerful and more efficient, head-mounted devices will continue to evolve, with more capabilities in "more wearable" devices. 

In terms of pure computing power, it is highly likely that we will see more ASICs (application-specific integrated circuits) inside those devices, that will be better suited for particular purposes such as computer vision (e.g. DSP for digital signal processor and VPU for vision processing unit).

Conventional silicon chips have already been pushed to their limits in terms of speed (or will soon be). In fact, recent years have focused on parallelizing several processors to try and compensate for the lack of speed increase. That being said, graphene processors are the future and will be much faster and smaller while requiring much less power \cite{graphene}. Those properties could obviously be helpful for wearable devices.

Similarly, computer vision and the SLAM problem will keep being active research domains. As algorithms get better, be it in terms of accuracy, robustness or complexity, spatial-aware AR experiences will improve. 

All those elements will probably lead to more broadly-available and affordable head-worn devices capable of enabling compelling AR/VR experiences but the upcoming addition of depth cameras to mobile phones will most likely play an even more important role in the democratization of spatial-aware AR. In fact, devices equipped with such depth sensors start to appear, e.g. with the ZenFone AR\footnote{\url{https://www.asus.com/Phone/ZenFone-AR-ZS571KL/}}, a high-end smartphone oriented towards consumers and that should be released in summer 2017.

\section{Interaction} \label{limitations-interaction}

Wearing an HMD enabling immersive or world anchored 3D experiences also requires special considerations with regards to how users can interact with it.

\begin{figure}[h]
	\includegraphics[width=7cm]{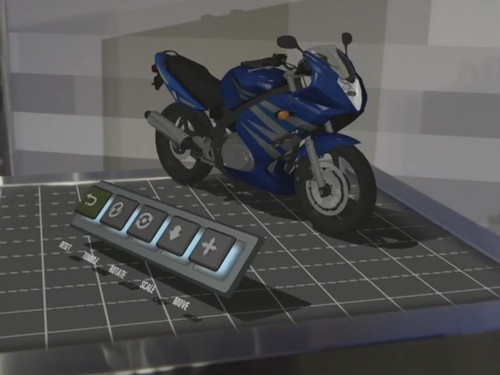}
	\centering
	\caption[Example of a UI anchored in the environment,  with buttons to interact with a 3D model]{\label{hololens-ui} Example of a UI anchored in the environment,  with buttons to interact with a 3D model\protect\footnotemark }
\end{figure}

Firstly, new kinds of UIs (user interfaces) need to be designed. In standard computer 3D applications such as games, information is often overlaid onto the virtual camera's view (e.g. to display a health bar). The same concept cannot be applied to optical see-through devices such as the Hololens, as the rendering's proximity to the user's eyes would be uncomfortable. The usual solution is to use a 3D UI integrated into the environment (possibly positioned depending on the user's gaze direction) instead of being attached to the camera. An example of such a UI is shown in figure \ref{hololens-ui}.

\footnotetext{\url{https://www.windowscentral.com/sites/wpcentral.com/files/styles/larger/public/field/image/2015/07/motorcycle-hololens.jpg?itok=gqeREJqm}} 

In order to interact with those UIs and other models, it is also necessary to provide some kind of gestural interaction as traditional input devices such as mouses and keyboards are generally not desirable (they are not adapted for that purpose). Speech recognition could also be an important part of those new interfaces but it still is a difficult problem, especially in noisy environments.

\begin{figure}[h]
	\includegraphics[width=10cm]{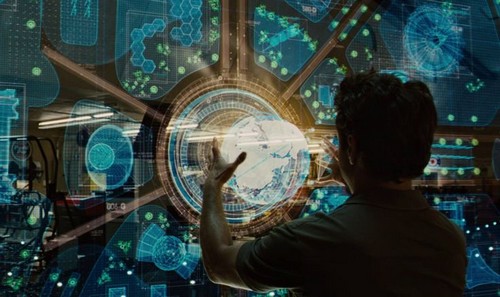}
	\centering
	\caption[Iron man's futuristic AR interface]{\label{ironman} Iron man's futuristic AR interface\protect\footnotemark }
\end{figure}

On a more personal note, I do believe that the usual approach of trying to "map" traditional input to a new kind of input (e.g. replacing a mouse click with a click gesture) might be the best approach for short term goals but is not advisable for the long run. New kinds of interactive environments should go along with new paradigms, even though technology is not ready yet for the futuristic AR interfaces (picture shown in figure \ref{ironman}) seen in the previously mentioned Iron Man movie\footnote{\url{https://www.youtube.com/watch?v=mRi1dmFgRfo}}.

\footnotetext{\url{https://s-media-cache-ak0.pinimg.com/736x/2f/49/84/2f4984329848be3825c17672beef797e.jpg}}

\begin{figure}[h]
	\includegraphics[width=10cm]{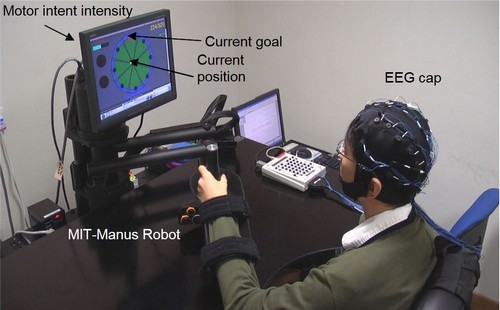}
	\centering
	\caption[A EEG-based BCI example with robotic feedback]{\label{bci-illustration} A EEG-based BCI example with robotic feedback \cite{bci-eeg}}
\end{figure}

Another way to get user input is by using brain-computer interfaces (BCIs). The idea is to analyze neural activity and map it to some kind of basic action. 
Different types of techniques can be used to monitor neural activity, invasive (implanted directly into the grey matter) and non-invasive (external devices) methods exist, with measurements typically taken by EEGs (electroencephalograms, analyzing the brain's activity) or EMGs (electromyogramsn analyzing muscular activity). Figure \ref{bci-illustration} pictures a non-intrusive system used in the context of neurorehabilitation in stroke.

Interesting results have been achieved, with promising applications, especially for individuals with muscular handicaps \cite{bci-disabled, bci-disabled2, bci-als} but also for certain types of autism, with BCI sometimes coupled with VR \cite{bci-vr-autism}.

However, current state-of-the-art methods are limited to a set of predefined simple actions and requires a significant amount of preliminary user training, mainly in research settings. Understanding complex intent (e.g. "go back to main menu") is completely out of reach as of now.

So far, this section only talked about what the user sees and how to get input from him but another important contribution will most likely come from haptic feedback. While VR enables immersive experiences, it is generally limited to viewing and listening to virtual content, which means only 2 out of 5 traditional senses can be used by the computer to send information to the user.
In those experiences, the users sometimes manipulate virtual objects but they usually cannot physically touch them. Haptic interfaces can be used to make users feel those virtual objects, by applying appropriate forces where needed.

A popular haptic interface is PHANTOM \cite{phantom}, developed by \citeauthor{phantom} in \citeyear{phantom}. Since then, the impact of such a technology has then been analyzed \cite{phantom-evaluation}, so was its complementarity with VR \cite{haptic-vr}. Commercial products also start to appear with (among many others) haptic gloves from ManusVR\footnote{\url{https://manus-vr.com/}} and even lots of complete body suits being in development such as the Teslasuit\footnote{\url{https://teslasuit.io/}} or Hardlight's suit\footnote{\url{http://www.hardlightvr.com/}}.

\section{Social acceptance}
Judging by the criticism around Google Glass\footnote{\url{https://www.google.com/glass/start/}} when it was released, mainly related to privacy issues, it looks like the general public is not ready to accept other people wearing cameras most of the time. While the HMDs discussed in chapter \ref{mr} dodged the issue by focusing on industrial applications, the problem remains for ordinary individuals.

On top of that, AR/VR wearables usually look futuristic but cannot really be considered good-looking by most people. While this is not crucial for their ability to deliver functional experiences, it might be important for the general public to embrace them.

Those concerns need to be addressed if we ever want to see AR wearables integrated into our daily lives, and the miniaturization of those devices will certainly help with that.

\chapter{Conclusion}

\label{conclusion}

I certainly learned a lot by completing this master's thesis, in terms of technologies and potential applications for AR and VR, two fields I am very interested in. I truly believe both AR and VR will be very beneficial to various industries in a relatively short term but exploring the possibilities also raised my expectations for our future daily lives.

Receiving the chance to work with a device as innovative as the Hololens was really attractive and the trouble related to the issues encountered while developing the game are no match to the joy of presenting it and seeing smiles on the players' face. I cannot thank the Microsoft Innovation Center enough for that opportunity.

I wish (and in fact plan on) extending the game's capabilities as real world geometry and objects could potentially be used to decide where the level's roads should be placed, using some kind of spatial-aware procedural level generation.

As said before, AR and VR will keep growing and it is definitely exciting, provided that the necessary evolutions in human-computer interactions follow the same path. Work remains to be done on several aspects of those domains and I do hope researchers will achieve significant progress to allow ubiquitous computing (every time, everywhere) and a seamless blending between reality and virtuality. Maybe even by developing wearable devices capable of switching back and forth between AR and VR modes with adaptive transparency?


\appendix 




\printbibliography[heading=bibintoc]


\end{document}